\documentclass[Journal]{IEEEtran}
\IEEEoverridecommandlockouts
\usepackage{graphicx, tikz, cite}
\usepackage{verbatim}
\usepackage{amsmath}
\usepackage{makeidx}
\usepackage{amsthm}
\usepackage{amssymb}
\usepackage{epstopdf}
\usepackage{multirow}
\usepackage{array}
\usepackage{makecell}
\usepackage{color}
\usepackage{hyperref}
\usepackage{graphicx, psfrag, minibox, mathrsfs, bigints, stfloats, color, upgreek, gensymb, textcomp, euscript, calligra, xcolor, pict2e}
\date{}
\usepackage{rotating}
\usepackage{diagbox}
\usepackage{slashbox}
\usepackage{tikz}
\usetikzlibrary{shapes,arrows}
\usetikzlibrary{automata,chains}
\usepackage{textcomp}
\usepackage{algorithmicx}
\usepackage{algpseudocode}
\usepackage[ruled]{algorithm2e}
\usepackage{setspace}
\usepackage{float}
\usepackage[font=scriptsize]{caption}
\usepackage{subcaption}
\usepackage{enumerate}
\usepackage{eso-pic}
\definecolor{red}{rgb}{0.8,0.1,0.1}
\definecolor{blue}{rgb}{0,0,1}
\definecolor{green}{rgb}{.1,.6,.3}
\newcommand{\red}{\color{red}}
\newcommand{\blue}{\color{blue}}
\newcommand{\green}{\color{green}}
\makeatletter
\newcommand{\leqnomode}{\tagsleft@true}
\newcommand{\reqnomode}{\tagsleft@false}
\makeatother
\begin{document}
\title{On Adaptive Transmission for Distributed Detection in Energy Harvesting Wireless Sensor Networks with Limited Fusion Center Feedback}
\author{\IEEEauthorblockN{Ghazaleh Ardeshiri, Azadeh Vosoughi~\IEEEmembership{Senior Member,~IEEE}} \thanks{Parts of this work were presented in GlobalSIP 2018  \cite{Ardeshiri} and GLOBECOM 2019 \cite{Ard2}.}}
\maketitle
\begin{abstract}
We  consider a wireless sensor network, consisting of $N$ heterogeneous sensors and a fusion center (FC), tasked with solving a binary distributed detection problem. Sensors communicate directly with the FC over orthogonal fading channels.
%
%
Each sensor can harvest randomly arriving energy and store it in a battery.  Also, it knows its quantized channel state information (CSI), acquired via a limited feedback channel from the FC. We propose a transmit power control strategy such that the $J$-divergence based detection metric is maximized, subject to  an  average  transmit  power  per  sensor  constraint. The proposed strategy is  parametrized  in terms  of  the  channel  gain  quantization  thresholds  and  the scale  factors corresponding to the quantization intervals, to  strike a balance between the rates of energy harvesting and energy consumption for data transmission. This strategy allows each sensor to adapt its transmit power based on its battery state and its qunatized CSI.
%
%
Finding the optimal strategy requires solving a non-convex optimization problem that is not  differentiable with respect to the optimization variables. We propose near-optimal strategy based on hybrid search methods that have a low-computational complexity. 
%
\end{abstract}
\begin{IEEEkeywords}
power control, distributed detection, channel gain quantization, energy harvesting,  $J$-divergence.
\end{IEEEkeywords}
\section{Introduction}
In a conventional wireless sensor network (WSN), sensors  powered by non-rechargeable batteries 
are used to sense and collect data for various applications. 
The energy constraint imposed by the non-rechargeable batteries has inspired a rich body of research on developing signal processing and transmission strategies to achieve balance between network lifetime and performance.
%
Recently, the technology of harnessing energy from the renewable resources of energy in ambient environment 
has attracted attention of many researchers, as a promising solution to address the challenging energy constraint problem in WSNs. In particular, energy harvesting (EH)-powered sensors offer potential for transforming design and performance of 
WSNs tasked with detection or estimation of a signal source \cite{advances}. 
In practice, the energy arrival of ambient energy sources is intrinsically time-variant and often sporadic. 
To flatten the randomness of the energy arrival, the harvested energy is stored in a battery, to balance the energy arrival and the energy consumption.
Power/energy management in EH-enabled WSNs with finite size batteries is necessary, in order to balance the rates of energy harvesting and energy consumption for transmission. 
%
%
If the energy management policy is overly aggressive, 
sensors may stop functioning, due to energy outage.
On the other hand, if the policy is overly conservative, sensors may fail to utilize the excess energy, due to energy overflow, leading into a performance degradation.

\begin{figure}[!t]
  \centering 
  \hspace{-24mm}
  \scalebox{0.9}{
    \begin{picture}(180,180)
      { \scriptsize
        \put(15,0){\includegraphics[width=75mm]{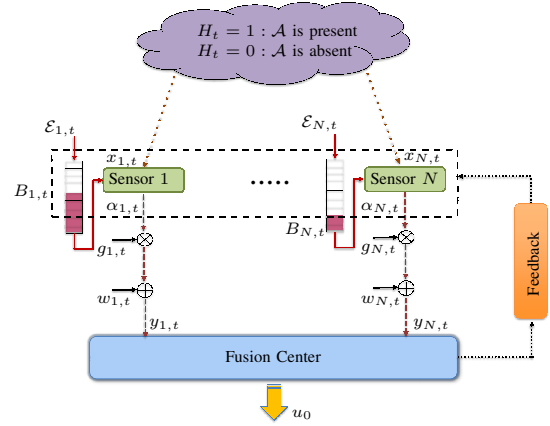}}
        \put(78,163){$H_t=1:  {\cal A}$ is present}   
        \put(78,155){$H_t=0:  {\cal A}$ is absent} 
        \put(90,26){Fusion Center}    
        \put(217,54){\rotatebox{90}{Feedback}}
        
        \put(41,101){Sensor $1$}    
        \put(40,110){$x_{1,t}$}
        \put(14,123){${\cal E}_{1,t}$}
        \put(1,96){$B_{1,t}$}
        \put(40,90){$\alpha_{1,t}$}
        \put(36,72){$g_{1,t}$}
        \put(36,51){$w_{1,t}$}
        \put(58,41){$y_{1,t}$}
        \put(100,101){{\Large .....}}

        \put(151,101){Sensor $N$}
        \put(165,111){$x_{N,t}$}
        \put(122,125){${\cal E}_{N,t}$}
        \put(115,80){$B_{N,t}$}
        \put(147,90){$\alpha_{N,t}$}
        \put(147,73){$g_{N,t}$}
        \put(147,51){$w_{N,t}$}
        \put(169,41){$y_{N,t}$}                

        \put(118,3){$u_0$}
      }
    \end{picture}
  }
\caption{Our system model and the schematic of battery state in time slot $t$.} 
\label{Fig-1-network}  
\end{figure}

In this work, we adopt a WSN
model that consists of several distributed sensors and a fusion center (FC).  The FC is tasked with solving a binary-hypothesis distributed detection problem. Each sensor is capable of harvesting energy from the ambient environment and is equipped with a battery of finite size to store the harvested energy. Sensors process locally their observations and communicate directly with the FC over orthogonal fading channels\footnote{The orthogonal channels are assigned using frequency-division duplexing.}. Each sensor only knows the quantized channel state information (CSI), via a limited feedback channel from the FC, and adapts its transmit power according to its battery state and its quantized CSI.  The FC jointly processes the received signals and makes a global decision  about the underlying hypotheses (see Fig. \ref{Fig-1-network}).  
%
%
In the following we provide a concise review of the literature  that are most related to our work. 
\subsection{Related Works and Knowledge Gap}
 The classical problem of binary distributed detection in a
network 
has a long and rich history  \cite{tenney1981,hoballah1989di}. 
%
%
%
Motivated by the potential applications of WSNs for event detection, researchers
have expanded these classical studies to include the effects of wireless communication channels between the sensors and the FC on the local decision rules at the sensors and the fusion rule at the FC \cite{Varshney2,ahmadi,maleki1}. 
To reduce energy consumption, researchers have further explored optimal power control strategies   \cite{vin,Goodman} that allow sensors to adapt their transmit powers based on the states of their propagation channels and the local sensor statistics.  In particular,  the authors in \cite{vin,Goodman} have designed the optimal power control strategies that maximize a $J$-divergence  based detection metric for  binary-hypothesis and multiple-hypothesis distributed detection problems, respectively. We note that \cite{vin,Goodman} assume that the CSI is perfectly available at the sensors for power control. 
%
%
%
%
However, CSI acquisition at the sensors in WSNs is difficult. In time division duplexing systems, sensors need to perform training-based channel estimation to acquire CSI  \cite{ahmadi1,ahmadi2,zahra}. In frequency division duplexing systems, sensors can acquire quantized CSI via a limited feedback channel from the FC \cite{Guo}. 
We note that signal adaptation at the sensors according to the quantized CSI received from a limited feedback channel has been considered before for data communications  \cite{marques} and distributed estimation of a signal source \cite{banavar,fanaei}.
%
%

It is worth pointing out that, while  the studies in \cite{vin,Goodman,ahmadi1,ahmadi2,zahra,Guo} on optimal power control strategies can be applied to WSNs with conventional battery-powered sensors, they cannot be applied to EH-enabled WSNs. None of these works have considered the new challenges related to power/energy management imposed by the random nature of the energy arrival and the harvested energy. 


In the context of distributed detection, there are only few studies that consider EH-powered sensors \cite{Tarighati,GengJun, Gupta}, among which \cite{Tarighati} is the closest work to ours. 
%
Modeling the battery state as a two-state Markov chain and choosing Bhattacharya distance as the detection performance metric, the authors in \cite{Tarighati} have investigated the optimal local decision thresholds at the sensors, such that the detection performance is optimized.
Considering an EH-powered node, that is deployed to monitor the change in its environment, the authors in \cite{GengJun} formulated a quickest change detection problem, where the goal is to detect the time at which the underlying distribution of sensor observation changes. 
%
Choosing error probability as the detection performance metric, the authors in \cite{Gupta} proposed ordered transmission schemes, that can lead to a smaller average
number of transmitting sensors, 
without comprising the detection performance. None of the works in \cite{Tarighati,GengJun, Gupta} have addressed transmit power control problem.  
Energy harvesting has been also considered in the contexts of cooperative data communication \cite{lioutage,wang}, 
distributed estimation of a signal source 
\cite{nourian,liu}, and 
cognitive radio systems \cite{taherpour, yazdani2020}. 



To the best of our knowledge, adaptive (channel-dependent) power control strategies in an EH-enabled WSN, where sensors can adapt their transmit powers based on their quantized CSI, with the goal of optimizing a detection metric, have not been explored. Hence, this is the focus of our work.  
\subsection{Our Contribution}
%
%

Given our adopted WSN model (see Fig. \ref{Fig-1-network}), we aim at developing a transmit power control strategy for sensors  that strikes a balance between energy harvesting and energy consumption for data transmission, and optimizes the detection performance. We choose the $J$-divergence between the distributions of the detection statistics at the FC under two hypotheses, as the detection performance
metric. Our choice is motivated by the facts that (i) it is a widely used metric for evaluating detection performance  \cite{vin,Guo,zahra,Goodman}, since it provides a lower bound on the detection error probability. Indeed, maximizing the $J$-divergence is equivalent to minimizing
the lower bound on the error probability; 
(ii) it allows us to provide a more
tractable analysis. Our proposed power control strategy is parametrized in terms of the channel gain quantization thresholds and
the scale factors (corresponding to the quantization intervals). The scale factors play key roles in balancing the rates of energy harvesting and  energy consumption for transmission.
We seek the jointly optimal scale factors and
and the quantization thresholds such 
that the $J$-divergence at the FC is maximized,  subject to an average transmit power per sensor constraint.  This optimization problem can be solved
{\it offline} at the FC, given the statistical information of fading
channels and the energy arrival. 
The solutions to this optimization problem is  available {\it a priori} at the sensors, such that each sensor can adapt its transmit power according to its battery state and its quantized CSI that is received from the FC via the feedback channel.  
%
%
Our main contributions can be summarized as follow:
 \begin{itemize}
     \item Our system model encompasses the stochastic energy arrival model for harvesting energy, and the stochastic energy storage model for the finite-size battery. We model the randomly arriving energy units during a time slot as a Poisson process, and the dynamics of the battery as a finite state Markov chain.
     \item We propose a novel parametrized power control strategy and formulate problem  (P1) to optimize the parameters such that the $J$-divergence at the FC is maximized, subject to an  average  transmit  power  per  sensor  constraint. 
     
    
     
     \item We derive an approximate expression for the detection error probability,
     relying on Lindeberg Central Limit Theorem (CLT) for large number of sensors.

     \item Since  (P1) is not concave with respect to the optimization variables,  and the objective function and the constraints in
     (P1) are not differentiable with respect to these variables, 
     we resort to grid-based search methods. In particular, we consider deterministic, random, and hybrid search methods, and explore the trade-offs in their performance and computational complexity. We show that the proposed hybrid search methods have the lowest computational complexity and provide a close-to-optimal performance. 
     
     \item We show that the optimized transmit power level is not a monotonic function of the channel gain (given the battery state), and explore the trade-off between transmit power  and  detection  performance.

     
 \end{itemize}
\subsection{Paper Organization}
The paper organization follows: Section \ref{sym} describes our system and observation models and 
introduces our constrained optimization problem (P1). Section \ref{J_error} derives a closed-form expression for the total $J$-divergence and an approximate expression for the error probability corresponding to the optimal Bayesian fusion rule at the FC. Sections \ref{cost_fun} and \ref{how-to-solve-P1} formulate and solve  problem (P1), respectively.  
 Section \ref{simulation} illustrates our numerical results. Section \ref{conclu} concludes our work. 
\section{System Model}\label{sym}
\subsection{Observation Model at Sensors}

%
%

To describe our signal processing blocks at sensors and the FC as well as energy harvesting model, we divide time horizon into slots of equal length $T_s$. Each time slot is indexed by an integer $t$ for $t=1,2,...,\infty$. We model the underlying binary hypothesis $H_t$ in time slot $t$ as a binary random variable $H_t \in \{0,1\}$ with a-priori probabilities $\Pi_0=\Pr(H_t=0)$ and $\Pi_1=\Pr(H_t=1)=1-\Pi_0$. We assume that the hypothesis $H_t$ varies over time slots in an independent and identically distributed (i.i.d.) manner. Let $x_{n,t}$ denote the local observation at sensor $n$ in time slot $t$. We assume that  sensors' observations given each hypothesis with conditional distribution $f(x_{n,t}|H_t=h_t)$ for $h_t \in \{0,1\}$ are independent across sensors. This model is relevant for WSNs that are tasked with detection of a known signal in uncorrelated Gaussian noises with the following signal model
\begin{align}\label{xk}
&H_{t}=1:~~ x_{n,t} ={\cal A}+v_{n,t},\nonumber\\
&H_{t}=0:~~ x_{n,t} = v_{n,t},~~
\text{for}~n=1,\dots,N,
\end{align}
where Gaussian observation noises $v_{n,t} \! \sim \! {\cal N}(0,\sigma_{v_{n}}^2)$ are independent over time slots and across sensors. 
Given observation $x_{n,t}$ sensor $n$ forms local log-likelihood ratio (LLR)
\begin{equation}\label{lrt_sensor}
 \Gamma_n(x_{n,t})\triangleq\log\left( \frac{f(x_{n,t}|h_{t}=1)}{f(x_{n,t}|h_{t}=0)}\right ),   
\end{equation}
and uses its value to choose its non-negative transmission symbol $\alpha_{n,t}$ to be sent to the FC.
In particular, when LLR is below a given local threshold $\theta_n$, sensor $n$ does not transmit and let $\alpha_{n,t}=0$. When LLR exceeds the given local threshold $\theta_n$, sensor $n$ chooses $\alpha_{n,t}$ according to its battery state and the feedback information about its communication channel. Choice of $\alpha_{n,t}$ will be explained later in Section \ref{battery model}.
\subsection{Battery State, Harvesting and Transmission Models
}\label{battery model}
We assume sensors are equipped with identical batteries of finite size $K$ cells (units), where each cell corresponds to $b_u$ Joules of stored energy. Therefore, each battery is capable of storing at most $K b_u$ Joules of harvested energy. 
Let $B_{n,t} \in \{0,1,...,K\}$ denote the discrete random process indicating the battery state of sensor $n$ 
at the beginning slot $t$. Note that $B_{n,t}=0$ and $B_{n,t}=K$ represent the empty battery and full battery levels, respectively. Also,  $B_{n,t}=k$ implies that the battery is at state $k$, i.e., $k$ cells of the battery is charged and the amount of stored
energy in the battery is $k b_u$ Joules. 

\par Let ${\cal E}_{n,t}$ denote the randomly arriving energy units\footnote{Suppose each arriving energy unit measured in Joules is $b_u$ Joules.} during time slot $t$ at sensor $n$. We assume ${\cal E}_{n,t}$'s are i.i.d. over time slots and across sensors. We model ${\cal E}_{n,t}$ as a Poisson random variable with parameter $\rho$, and probability mass function (pmf) $p_e \triangleq\Pr({\cal E}_{n,t}=e)= e^{\rho}\rho^e/e!$ for $e=0,1,\dots,\infty$.
Note that parameter $\rho$ is the average number of arriving energy units during one time slot at each sensor. Let ${\cal S}_{n,t}$ 
be the number of stored (harvested) energy units in the battery at sensor $n$ during time slot $t$. Note that the harvested energy ${\cal S}_{n,t}$ cannot be used during slot $t$. 
Since the battery has a finite capacity of $K$ cells, we have ${\cal S}_{n,t} \in \{0,1,...,K\}$. Also, ${\cal S}_{n,t}$ are i.i.d. over time slots and across sensors. We  can find the pmf of ${\cal S}_{n,t}$ in terms of the pmf of ${\cal E}_{n,t}$. Let $q_e\triangleq \Pr({\cal S}_{n,t}=e)$ for $e=0,1,\dots,K$. We have\footnote{ Equation \eqref{harvesting} assumes that the energy storage process is lossless. For a lossy storage process, one needs to model such loss via establishing a functional relationship between ${\cal S}_{n,t}$ and ${\cal E}_{n,t}$, i.e., ${\cal S}_{n,t} = f_n({\cal E}_{n,t})$, where the function $f_n(.)$ can be approximated using the battery type and specifications. Knowing $f_n(.)$ and the pmf of ${\cal E}_{n,t}$, one can find the pmf of ${\cal S}_{n,t}$ using transformation
method.}
\begin{equation}\label{harvesting}
    q_e=\begin{cases}
    p_e,~~~~~~~~~~~~~~\text{if}~ 0\leq e \leq K-1,\\
    \sum_{m=K}^\infty p_m,~~~~~\text{if}~ e=K.
    \end{cases}
\end{equation}
\par Let $g_{n,t}$ indicate the fading channel gain between sensor $n$ and the FC during time slot $t$. %
We assume block fading model and $g_{n,t}$'s  are  i.i.d.  over time slots and independent across sensors. 
%
%
We assume there is a limited feedback channel from the FC to the sensors \cite{Guo}, through which sensor $n$ is informed of the quantization interval to which $g_{n,t}$ belongs. 
In particular, suppose the positive real line is partitioned into $L$ disjoint intervals $\mathcal{I}_{n,l}= [\mu_{n,l},\mu_{n,l+1})$ for $l=0,...,L-1$,  using the quantization thresholds $\{ \mu_{n,l}\}_{ l=0}^L$, where $0\! = \mu_{n,0}\! < \mu_{n,1}\! < \dots \!< \mu_{n,L}\! = \infty$  (to be optimized). The quantization mapping rule follows: if the quantizer input $g_{n,t}$ lies in the interval  $\mathcal{I}_{n,l}$ then the quantizer output is $\mu_{n,l}$.
%
%
Let
$\pi_{n,l}=\Pr(g_{n,t}\in \mathcal{I}_{n,l})$ be the probability that $g_{n,t}$ lies in the interval  $\mathcal{I}_{n,l}$. This probability depends on the distribution of fading model.
%
For instance, for Rayleigh fading model $g_{n,t}^2$ has exponential distribution with the mean $\mathbb{E}\{g_{n,t}^2\}=\gamma_{g_n}$ and we have 
\begin{equation}\label{gaz}
\pi_{n,l}=\Pr\Big((g_{n,t}^2 \in [\mu_{n,l
}^2, \mu_{n,l+1}^2)\Big)= e^{\frac{-\mu_{n,l}^2}{\gamma_{g_n}}}-e^{\frac{-\mu_{n,l+1}^2}{\gamma_{g_n}}}.
\end{equation}
Let ${\cal P}_{n,t}$ denote 
the transmit power of sensor $n$ in time slot $t$. 
When LLR is below a given local threshold $\theta_n$, sensor $n$ does not transmit, i.e., ${\cal P}_{n,t}=0$. When LLR
exceeds $\theta_n$, 
sensor $n$ chooses ${\cal P}_{n,t}$ according to its battery state $k$ and the feedback information. In
particular, we choose a transmit power control strategy where ${\cal P}_{n,t}$ is proportional to the amount of stored energy in the battery, i.e., $kb_u$ Joules, and the scale factor depends on the feedback information. 
%
%
%
Mathematically, we express ${\cal P}_{n,t}$ as the following
\begin{equation}\label{alpha}
{\cal P}_{n,t} =
\begin{cases}
0,&~~~\Gamma_n(x_{n,t})<\theta_n,\\
\lfloor c_{n,0} k\rfloor b_u/T_s,&~~~\Gamma_n(x_{n,t})\geq\theta_n,~ g_{n,t}\in\mathcal{I}_{n,0},\\
\vdots&~~~~~~~~~~~~~\vdots\\
\lfloor c_{n,L-1} k\rfloor b_u/T_s ,&~~~\Gamma_n(x_{n,t})\geq\theta_n,~g_{n,t}\in\mathcal{I}_{n,L-1},
\end{cases}
\end{equation}
where $\lfloor . \rfloor$ is the floor function and the scale factors $\{c_{n,l}\}_{l=0}^{L-1}$ are between zero and one. The number of scale factors is equal to the number of quantization levels and scale factor $c_{n,l}$ corresponds to the quantization interval $\mathcal{I}_{n,l}=[\mu_{n,l},\mu_{n,l+1})$. 
Given $\theta_n$, the problem of optimizing transmit power control strategy reduces to finding the best scale factors $\{c_{n,l}\}_{l=0}^{L-1}$ and the quantization thresholds $\{\mu_{n,l}\}_{l=1}^{L-1}$ such that a specified performance metric is optimized. We let the transmit symbol   
$\alpha_{n,t}=\sqrt {{\cal P}_{n,t}}$.
Considering the power control strategy in (\ref{alpha}), we note that 
the number of energy units consumed for transmitting  symbol $\alpha_{n,t}$ is 
$\lfloor c_{n,l} k\rfloor$, which is an integer between zero and $K$ and is always smaller than $k$. In other words, the energy consumption for transmission cannot exceed the stored energy in the battery, and  
%
%
the battery cannot be fully depleted
after a transmission.  
It also implies that when $\lfloor c_{n,l} k\rfloor=0$ the sensor will not transmit. 
%
%
Note that the scale factors $\{c_{n,l}\}_{l=0}^{L-1}$ in \eqref{alpha} play key roles in balancing the rates of energy harvesting and energy consumption for transmission. 
Given the quantization thresholds $\mu_{n,l}$'s, when $c_{n,l}$'s
are closer to one, such that the rate of energy consumption for transmission is greater than the rate of energy harvesting, sensors may stop functioning, due to energy outage. When $c_{n,l}$'s
are closer to zero, such that the rate of energy consumption for transmission is smaller than the rate of energy harvesting, sensors may fail to utilize the excess
energy, due to energy overflow, leading into a performance degradation. 
%

\begin{figure}[!t]
  \centering
  \hspace{-0mm}
  \scalebox{0.9}{
    \begin{tikzpicture}[start chain=going left,node distance=1.3cm]   
      \scriptsize
      \node[state, on chain]                 (K) {$K$};
      \node[state, on chain]                 (K-1) {$\!K\!\!-\!\!1\!$};
      \node[on chain]                   (g) {$\cdots$};
      \node[state, on chain]                 (1) {$1$};
      \node[state, on chain]                 (0) {$0$};
      \draw[
      >=latex,
      auto=right,                      
      loop above/.style={out=75,in=105,loop},
      every loop,
      ]
      (0)   edge[loop above] node {$\psi_{0,0}$}   (0)
      edge             node {$\psi_{0,1}$} (1)
      (1)   edge[loop above] node {$\psi_{1,1}$}   (1)
      edge             node {$\psi_{1,2}$}   (g)
      (g)  edge[] node {$\psi_{K\!-\!2,K\!-\!1}$}   (K-1)
      (K-1)   edge[loop above] node {$\psi_{K\!-\!1,K\!-\!1}$}   (K-1)
      edge             node {$\psi_{K\!-\!1,K}$}   (K)
      (K-1)   edge[bend left] node[below] {$\psi_{K\!-\!1,0}$}   (0)
      (K)   edge[loop above] node {$\psi_{K,K}$}   (K-1)
      (K)   edge[bend left] node[below] {$\psi_{K,1}$}   (1);
    \end{tikzpicture}
  }
  \caption{ Schematics of Markov chain corresponding to the battery state random process $B_{n,t}$.} 
  \label{fig_battery}
\end{figure}
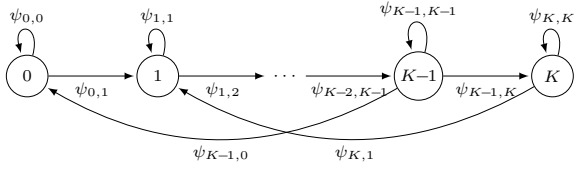

The battery state at the beginning of slot $t+1$ depends on the battery state at the beginning of slot $t$, the harvested energy during slot $t$, and the number of stored energy units that is consumed for transmitting symbol $\alpha_{n,t}$, i.e., ${\cal P}_{n,t} T_s/b_u$. Mathematically, we express $B_{n,t+1}$ as the following 
\begin{equation}\label{b_n,t}
    B_{n,t+1} = \min\big\{[B_{n,t} + {\cal S}_{n,t}-{\cal P}_{n,t} T_s/b_u]^+,K \big \},
\end{equation}
where $[x]^+=\max \{0,x\}$. Considering the dynamic battery state model in \eqref{b_n,t} we note that, conditioned on ${\cal S}_{n,t}$ and ${\cal P}_{n,t}$ the value of $B_{n,t+1}$ only depends on the value of $B_{n,t}$ (and not the battery states of time slots before $t$). Hence, the process $B_{n,t}$ can be modeled as a Markov chain. Fig. \ref{fig_battery} is the schematic representation of this $(K+1)$-state Markov chain.
Let $\boldsymbol{\Phi}_{n,t}$ be the probability vector of battery state in slot $t$
\begin{equation}\label{battery_state}
  \boldsymbol{\Phi}_{n,t}\triangleq\Big[\Pr(B_{n,t}=0),\dots,\Pr(B_{n,t}=K)\Big]^T,
\end{equation}
where the superscript $T$ indicates transposition. We note that $\Pr(B_{n,t}=k)$ in \eqref{battery_state} depends on $B_{n,t-1}$, ${\cal S}_{n,t-1}$ and ${\cal P}_{n,t-1}$. Assuming that the Markov chain is time-homogeneous\footnote{ A Markov chain is time-homogeneous (stationary) if and only if its
transition probability matrix is time-invariant. Adopting homogeneous Markov
chain model for studying EH-enabled communication systems is widely
common \cite{short}.}, we let $\boldsymbol{\Psi}_n$ be the corresponding $(K+1) \times (K+1)$ transition probability matrix of this chain with its $(i,j)$-th entry 
$\psi_{i,j}\triangleq\Pr(B_{n,t}=j|B_{n,t-1}=i)$ for $i,j = 0,\dots, K$.
 Defining the indicator function $I_{i\rightarrow j}({\cal S}_{n,t},{\cal P}_{n,t}T_s/b_u)$ as \eqref{indicator}.
\begin{figure*}
\begin{equation}\label{indicator}
    I_{i\rightarrow j}({\cal S}_{n,t},{\cal P}_{n,t}T_s/b_u)=
    \begin{cases}
    1,&\text{if}~ j\!=\!\min \big\{[i+{\cal S}_{n,t}-{\cal P}_{n,t}T_s/b_u]^+,K\big\},\\
    0,&\text{o.w.}
    \end{cases}
\end{equation}
 \hrulefill
\end{figure*}
We can express $\psi_{i,j}$ as below

%
%
\begin{align}
    \psi_{i,j}\!=\! \widehat{\Pi}_{n,1}\sum_{k=0}^K \sum_{l=0}^{L}\pi_{n,l}q_k I_{i\rightarrow j}({\cal S}_{n,t},\lfloor c_{n,l} i\rfloor)\nonumber
\end{align}    
\begin{align}\label{pij}
    & \!\!\!\!\!\!\!\!\!\!\!\!\!+\!\widehat{\Pi}_{n,0}\!\sum_{k=0}^K q_k I_{i\rightarrow j}({\cal S}_{n,t},0).
\end{align}
%
The symbols $\widehat{\Pi}_{n,0}$ and $\widehat{\Pi}_{n,1}$ in \eqref{pij} refer to the probabilities of events ${\cal P}_{n,t} = 0$ and ${\cal P}_{n,t} \neq 0$, respectively. In particular, we have
\begin{align}
    &\widehat{\Pi}_{n,0} = \Pr({\cal P}_{n,t}\! =\! 0) = \Pi_0(1\!-\!P_{\text{f}_n}) + \Pi_1(1\!-\! P_{\text{d}_n}),\nonumber\\
    &\widehat{\Pi}_{n,1}= \Pr({\cal P}_{n,t}\! \neq \! 0) = \Pi_0 P_{\text{f}_n} + \Pi_1 P_{\text{d}_n},
\end{align}
where the probabilities $P_{\text{f}_n}$ and $P_{\text{d}_n}$ can be determined using our signal model in \eqref{xk}
\vspace{-2mm}
\begin{align}\label{pd_pf1}
 &P_{\text{f}_n} \!= \! \Pr({\cal P}_{n,t}\!\neq\!0 |h_t=0)\!=\!Q\Big(\frac{\theta_n+{\mathcal{A}^2}/{2\sigma^2_{v_n}}}{\sqrt{\mathcal{A}^2/{\sigma^2_{v_n}}}}\Big), \nonumber\\
 &P_{\text{d}_n} \!= \!\Pr({\cal P}_{n,t}\!\neq \!0 |h_t=1)\!=\!Q\Big(\frac{\theta_n-{\mathcal{A}^2}/{2\sigma^2_{v_n}}}{\sqrt{\mathcal{A}^2/{\sigma^2_{v_n}}}}\Big). 
\end{align}
Suppose $P_{\text{d}_n}$ is required to be fixed at a given value  $P_{\text{d}_n} = \overline{P}_{\text{d}}, \forall n$.  Then the false alarm probability can be written as
    $P_{\text{f}_n} = Q\left(Q^{-1}(\overline{P}_{\text{d}})+\sqrt{\mathcal{A}^2/{\sigma^2_{v_n}}}\right)$.
Going back to the transition probability matrix $\boldsymbol{\Psi}_n$, since the Markov chain characterized by  $\boldsymbol{\Psi}_n$ is irreducible and aperiodic, there exists a unique steady state  distribution, regardless of the initial state \cite{short}. Let $\boldsymbol{\Phi}_n=[\phi_{n,0}, \phi_{n,1}, ...,\phi_{n,K}]^T$ be the unique steady state probability vector with the entries $\phi_{n,k}=\lim _{t\rightarrow \infty} \Pr(B_{n,t}=k)$. Note that this vector satisfies the following eigenvalue equation 
\begin{equation}\label{inf}
 \boldsymbol{\Phi}_{n}= \boldsymbol{\Phi}_{n}\boldsymbol{\Psi}_n.
\end{equation}
In particular, we let $\boldsymbol{\Phi}_{n}$ be the normalized eigenvector of $\boldsymbol{\Psi}_n$ corresponding to the unit eigenvalue, such that the sum of its entries is one \cite{Tarighati}. The closed-form expression for $ \boldsymbol{\Phi}_{n}$ can be written as \cite{yazdani2020}
\begin{equation}\label{inverse_battery}
 \boldsymbol{\Phi}_{n} = -(\boldsymbol{\Psi}_n^T-\textbf{I}-\textbf{B})^{-1}\textbf{1},
\end{equation}
where $\textbf{B}$ is an all-ones matrix, $\textbf{I}$ is the identity matrix, and \textbf{1} is an all-ones column vector. 
From this point forward, we assume that the battery operates at its steady state and we drop the superscript $t$.

\begin{figure}[!t]
\begin{subfigure}[t]{0.5\textwidth}
  \centering
  \includegraphics[scale=.47]{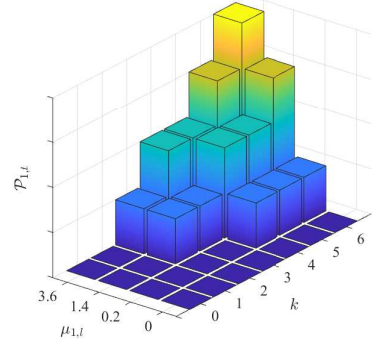}  
  \caption{ $\mu_{1}^{(a)}=[0,0.2,1.4,3.6],$ $c_{1}^{(a)}=[0.1,0.3,0.5,0.7]$}
 \label{powermap1}
\end{subfigure}
\begin{subfigure}[t]{0.5\textwidth}
  \centering
  \includegraphics[scale=.47]{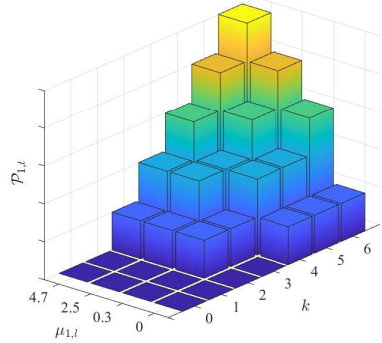}  
  \caption{ $\mu_{1}^{(b)}=[0,0.3,2.5,4.7],$ $c_{1}^{(b)}=[0.3,0.5,0.7,0.9]$}
  \label{powermap2}
\end{subfigure}
\caption{This example shows how much power ${\cal P}_{1,t}$ the single sensor  should spend for its data transmission, given  its  battery state and the feedback information.}
\label{alpha_map}
\end{figure}

For clarity of the presentation and to illustrate our transmit power control strategy in \eqref{alpha}, we consider the following simple example consisting of one sensor, i.e., $N\!=\!1$, and let $L\!=\!4,~K\!=\!6,~\rho \!=\! 2$ and $\gamma_{g_1}=1$. To examine the effect of variations of the scale factors and the quantization thresholds on $\boldsymbol{\Psi}_{n}$ and $\boldsymbol{\Phi}_n$ and transmit power, we consider two sets of values  $c_{1}^{(a)}=[0.1,~0.3,~0.5,~0.7]$, $\mu_{1}^{(a)}=[0,~0.2,~1.4,~3.6,~\infty]$ and $c_{1}^{(b)}=[0.3,~0.5,~0.7,~0.9]$, $\mu_{1}^{(b)}=[0,~0.3,~2.5,~4.7,~\infty]$. The corresponding $7 \times 7$ transition matrices, denoted as $\boldsymbol{\Psi}_1^{(a)}$ and $\boldsymbol{\Psi}_1^{(b)}$, as well as the corresponding $ 7 \times 1$ steady state probability vectors, denoted as $\boldsymbol{\Phi}_1^{(a)}$ and $\boldsymbol{\Phi}_1^{(b)}$ are
\vspace{-2mm}
\begin{align*}\label{psi1}
\boldsymbol{\Psi}_1^{(a)} = 
\small
\begin{pmatrix}
0.13& 0.27&	0.27& 0.17&	0.09& 0.03&	0.04\\   
0&	0.13& 0.27&	0.27& 0.18& 0.09& 0.06\\     
0&	0.02&	0.15&	0.27&	0.25&	0.16&	0.15\\
0&	0&	0.02&	0.15&	0.27&	0.26&	0.30\\ 
0&	0&	0.02&	0.09&	0.21&	0.25&	0.43\\
0&	0&	0&	0.02&	0.09&	0.21&	0.68\\
0&	0&	0&	0.02&	0.04&	0.09&	0.85&\\
\end{pmatrix},\nonumber
\\\boldsymbol{\Psi}_1^{(b)}=
\small
\begin{pmatrix}
0.14 & 0.28& 0.28& 0.16& 0.07& 0.04&	0.03\\ 
0 & 0.14& 0.28&	0.28& 0.16& 0.09& 0.05\\ 
0&	0.06&	0.19&	0.27&	0.22&  0.13& 0.13\\
0&	0&	0.07&	0.20&	0.27&  0.22& 0.24\\  
0&	0&	0.06&	0.15&	0.22&	0.22&	0.35\\
0&	0&	0&	0.07&	0.15&	0.21&	0.57\\
0&	0&	0&	0.07&	0.12&	0.14&	0.67&\\
\end{pmatrix}.
\end{align*}
\begin{align*}
    &\boldsymbol{\Phi}_{n}^{(a)}=[ 0, 0.0004,  0.0027, 0.0290, 0.0640, 0.1195,  0.7844]\\
    &\boldsymbol{\Phi}_{n}^{(b)}=[0,
    0.0015, 0.0209,0.1002, 0.1582, 0.1723,  0.5469]
\end{align*}
Given these two sets of values,  Fig. \ref{alpha_map} illustrates the two corresponding transmit power maps  assuming $b_u \!=\! 10$ mJ and $T_s \!=\! 10$ sec. The transmit power maps in \eqref{alpha} show how much power  the sensor should spend for its data transmission, given its battery state $k$ and the feedback information (i.e., the quantization interval to which the channel gain $g_{1,t}$ belongs). 
%
 For instance, for the parameters in Fig. (\ref{powermap1}), when $g_{1,t} \in \mathcal{I}_{1,2}$ and $B_{1,t}=3$, then ${\cal P}_{1,t}=1 $ mW, whereas for the parameters in Fig. \ref{powermap2}, when $g_{1,t} \in \mathcal{I}_{1,2}$ and $B_{1,t}=3$, then ${\cal P}_{1,t}=2$ mW.
 \subsection{Received Signals at FC and Optimal Bayesian Fusion Rule}
In each time slot sensors send their data symbols to the FC over orthogonal fading channels.   
 The received signal at the FC from sensor $n$ corresponding to time slot $t$ is
\begin{equation}\label{y_n,t}
   y_{n,t} = g_{n,t}\, \alpha_{n,t} + w_{n,t}, ~~~~\text{for}~n=1,\dots,N
\vspace{-1mm}
\end{equation}
where $w_{n,t} \sim {\cal N} (0,\sigma_{w_n}^2)$ is the additive Gaussian noise and $\alpha_{n,t}=\sqrt {{\cal P}_{n,t}}$. We assume $w_{n,t}$'s are i.i.d. over time slots and independent across sensors. 
 Let $\boldsymbol{y}_t=[y_{1,t},y_{2,t}, \ldots ,y_{N,t}]$ denote the vector that includes the received signals at the FC from all sensors in time slot $t$. The FC applies the optimal Bayesian fusion rule $\Gamma_0(.)$ to the received vector $\boldsymbol{y}_t$ and obtains a global decision $u_{0,t}=\Gamma_0(\boldsymbol{y}_t)$, where $u_{0,t} \in \{0,1\}$ \cite{ahmadi}. In particular, we have
 \begin{equation}\label{u0_lrt}
 u_{0,t}= \Gamma_0(\boldsymbol{y}_t)= \\
 \begin{cases}
 1,~~~~~~\Delta_t > \tau, \\
 0,~~~~~~ \Delta_t < \tau,
 \end{cases}
\end{equation}
 where the decision threshold $\tau =\log( \frac{\Pi_0}{\Pi_1})$ and 
\begin{equation}\label{lrt1}
\Delta_t=\log\left (\frac{f(\boldsymbol{y}_t|h_t=1)}{f(\boldsymbol{y}_t|h_t=0)}\right),
\end{equation}
and $f(\boldsymbol{y}_t|h_t)$ is the conditional probability density function (pdf) of the received vector $\boldsymbol{y}_t$ at the FC. 
\subsection{Our Proposed Constrained Optimization Problem}
From Bayesian perspective, the natural choice to measure the detection performance corresponding to the global decision $u_{0,t}$ at the FC is the error probability, defined as
\begin{align}\label{pe1}
 P_e &=\Pi_0 \Pr (u_{0,t}=1|h_t=0)+\Pi_1 \Pr(u_{0,t}=0|h_t=1)\nonumber\\
&=\Pi_0 \Pr(\Delta_t > \tau |  h_t=0) + \Pi_1 \Pr(\Delta_t < \tau |  h_t=1).   
\end{align}
 However, finding a closed form expression for $P_e$ is often mathematically intractable. 
Instead, we choose the total $J$-divergence between the distributions of the detection statistics at the FC under different hypotheses (will be defined in  Section \ref{J_error}), as our detection performance metric. This choice allows us to provide a more tractable analysis. 

Our goal is to find the scale factors 
$\{c_{n,l}\}_{l=0}^{L-1}$ 
and the quantization thresholds 
$\{\mu_{n,l}\}_{l=1}^{L-1}$
in the transmit power control strategy  (\ref{alpha})  for all sensors such that the total $J$-divergence at the FC is maximized,  subject to an average transmit power per sensor constraint. We assume that this optimization problem is solved
{\it offline} at the FC, given (i) the statistical information of fading
channels and noises (including communication channel noise and observation noise) and randomly arriving energy units, and (ii) the battery  parameter $K$, the number of quantization levels $L$, and the given 
$\overline{P}_{\text{d}}$ for local detectors at the sensors. 
The solutions to this optimization problem is available a {\it a priori} at the FC
and the sensors, to be utilized for controlling and adapting transmit power according to (\ref{alpha}). The idea of offline power control optimization with a limited feedback channel has been used
before for distributed detection systems in WSNs \cite{Guo}.
%

\section{Characterization of Total \textit{J}-divergence and Error Probability}\label{J_error}
In this section, first we define the total $J$-divergence and then derive a closed-form expression for it in Section III.A, using Gaussian distribution approximation. Next, considering 
$P_e$ in \eqref{pe1}  we provide a closed-form approximate expression for it in Section III.B,  using the same Gaussian distribution approximation and Lindeberg central limit theorem (CLT). 

\subsection{Total \textit{J}-Divergence Derivation}\label{j_drive}
We start with the definition of $J$-divergence. Consider two pdfs of a continuous random variable $x$, denoted as $\eta_1(x)$ and $\eta_2(x)$. By definition \cite{vin}, \cite{zahra}, the $J$-divergence between $\eta_1(x)$ and $\eta_0(x)$, denoted as $J(\eta_1,\eta_0)$, is
\begin{equation}\label{first_j}
    J(\eta_1,\eta_0) = D(\eta_1||\eta_0)+D(\eta_0||\eta_1),
\end{equation}
where $D(\eta_i||\eta_j)$ is the non-symmetric Kullback-Leibler (KL) distance between $\eta_i(x)$ and $\eta_j(x)$. The KL distance $D(\eta_i||\eta_j)$ is defined as
\begin{equation}\label{kl}
    D(\eta_i||\eta_j) = \int_{-\infty}^{\infty} \log \left(\frac{\eta_i(x)}{\eta_j(x)}\right)\eta_i(x) dx.
\end{equation}
Substituting \eqref{kl} into \eqref{first_j} we obtain
\begin{equation}\label{j23}
J(\eta_1,\eta_0)=\int_{-\infty}^{\infty} \left[\eta_1 (x)-\eta_0 (x)\right] \log \left(\frac{\eta_1(x)}{\eta_0(x)}\right) dx. 
\end{equation}
In our problem setup, the two conditional pdfs $f(\boldsymbol{y} |h=1)$ and $f(\boldsymbol{y} |h=0)$ play the role of $\eta_1(x)$ and $\eta_0(x)$, respectively. Let $J_{tot}$ denote the $J$-divergence between $f(\boldsymbol{y}| h=1)$ and  $f(\boldsymbol{y}|h=0)$. The pdf of vector $\boldsymbol{y}$ given $h$ is
\begin{align}\label{y/H}
f(\boldsymbol{y}|h)&\overset{(a)}=\prod_{n=1}^N f(y_n|h)   \nonumber\\
& \overset{(b)}= \prod_{n=1}^N f(y_{n}|\alpha_{n}, h)\Pr(\alpha_{n}|h)
\nonumber\\&\overset{(c)}=\prod_{n=1}^N \underbrace{f(y_{n}|\alpha_{n})\Pr(\alpha_{n}|h)}_{=f(y_{n}|h)}, ~~~~~~\text{for}~h=0,1.
\end{align}
Equality ($a$) in \eqref{y/H} holds since the received signals from sensors at the FC, given $h$, are conditionally independent, equality ($b$) in \eqref{y/H} is obtained from Bayes' rule, and equality ($c$) in \eqref{y/H} is found noting that $H$, $\alpha_n$, $y_n$ satisfy the Markov property, i.e.,  $H \rightarrow \alpha_n \rightarrow y_n$ \cite{vin}, \cite{zahra} and hence $y_n$ and $H$, given $\alpha_n$, are conditionally independent. 
Let $J_n$ represent the $J$-divergence between the two conditional pdfs $f({y}_n|h=1)$ and $f({y}_n|h=0)$. Using \eqref{j23} we can express $J_n$ as
\begin{align}
    J_n = &\nonumber\\
   \int_{-\infty}^{\infty}& \!\!\Big [f(y_n| h=1)\!-\! f(y_n| h=0)\Big ] {\rm log}\left({\frac{f(y_n| h=1)} {f(y_n|h=0)}}\right)\;dy_n.
\end{align}
Based on \eqref{y/H} we have $J_{tot}= \sum_{n=1}^N J_n.$
To calculate $J_n$, we need to find the conditional pdf $f(y_n|h)$. Considering \eqref{y_n,t} we realize that $y_n$, given $\alpha_n$, is Gaussian.
In particular, we have
\begin{equation}\label{Gaussian-dis}
f(y_n | \alpha_n=0)= \mathcal{N}(0, \sigma_{w_n}^2), f(y_n |\alpha_n \neq 0)=\mathcal{N}(g_n \alpha_n, \sigma_{w_n}^2)
\end{equation}
Also, considering \eqref{pd_pf1} and noting that $\alpha_{n}=\sqrt {{\cal P}_{n}}$
we find
\begin{align}\label{Gaussian-dis2}
  &\Pr(\alpha_n\neq 0|h=0)= P_{\text{f}_n},~
  \Pr(\alpha_n \neq 0 |h=1)= P_{\text{d}_n},\\
  &\Pr(\alpha_n = 0 |h=0) = 1\!-\!P_{\text{f}_n},~ \Pr(\alpha_n = 0|h=1) = 1\!-\! P_{\text{d}_n}.\nonumber  
\end{align}
Substituting \eqref{Gaussian-dis} and \eqref{Gaussian-dis2} in \eqref{y/H}, the conditional pdfs $f(y_n|h=0)$ and $f(y_n|h=1)$ become
\begin{align}
\label{xyz}
&f(y_n |h=0)=f(y_n |\alpha_n \neq 0) P_{\text{f}_n} + f(y_n |\alpha_n = 0) (1-P_{\text{f}_n}),\nonumber\\
&f(y_n |h=1)=f(y_n |\alpha_n \neq 0) P_{\text{d}_n} + f(y_n |\alpha_n = 0) (1-P_{\text{d}_n}).
\end{align}
Although $f(y_n |\alpha_n=0)$ and $f(y_n | \alpha_n \neq 0)$ in \eqref{xyz} are Gaussian, $f(y_n|h=0)$ and $f(y_n|h=1)$ are Gaussian mixtures, due to $P_{\text{d}_n}$ and $P_{\text{f}_n}$. 
Unfortunately, the $J$-divergence between two Gaussian mixture densities does not have a general closed-form expression. Similar to \cite{vin}, \cite{zahra} we approximate the $J$-divergence between two Gaussian mixture densities by the $J$-divergence between two Gaussian densities $f^G(y_n|h) \sim {\cal N}(m_{n,h},\Upsilon_{n,h}^2)$, where the mean $m_{n,h}$ and the variance $\Upsilon_{n,h}^2$ of the approximate distributions are obtained from matching the first and second order moments of the actual and the approximate distributions. For our problem setup, one can verify that the parameters $m_{n,h}$ and $\Upsilon_{n,{h}}^2$ become
\begin{align}\label{G_par}
    &m_{n,0}={{g_n}\alpha_n}P_{\text{f}_n}\nonumber,~~~ \Upsilon_{n,0}^2\!=\!g_n^2\alpha_n^2 P_{\text{f}_n}(1\!-\!P_{\text{f}_n})\!+\!\sigma_{w_n}^2,\\ 
&m_{n,1}\!=\!{{g_n}\alpha_n}P_{\text{d}_n},~~~~ \Upsilon_{n,1}^2\!=\!g_n^2\alpha_n^2 P_{\text{d}_n}(1\!-\!P_{\text{d}_n})\!+\!\sigma_{w_n}^2.
\end{align}
The $J$-divergence between two Gaussian densities, represented as $J\big(f^G(y_n|h=1),f^G(y_n|h=0)\big)$, in terms of their means and variances is \cite{vin}
\begin{align}\label{g_J}
    & J\big(f^G(y_n|h=1),f^G(y_n| h=0)\big)= \nonumber\\
     &\frac{\Upsilon_{n,1}^2\!+\!( m_{n,1}\!-\! m_{n,0})^2}{\Upsilon_{n,0}^2}
     +\frac{\Upsilon_{n,0}^2\!+\!( m_{n,0}\!-\! m_{n,1})^2}{\Upsilon_{n,1}^2}. 
\end{align}
Substituting $m_{n,h}$ and $\Upsilon_{n,h}^2$ into $J_n$ in \eqref{g_J} we approximate $J_n$ as the following\vspace{-2mm}
\begin{align}\label{sim_j}
    J_n=\frac{\sigma_{w_n}^2 + A_n g_n^2\alpha_n^2}{\sigma_{w_n}^2 + B_n g_n^2\alpha_n^2}&+\frac{\sigma_{w_n}^2  + C_n g_n^2\alpha_n^2}{\sigma_{w_n}^2  +  D_n g_n^2\alpha_n^2},
\end{align}
 where
\begin{align*} 
A_n =&~P_{\text{f}_n}(1\!-\!P_{\text{d}_n}) + P_{\text{d}_n}(P_{\text{d}_n}\!-\!P_{\text{f}_n}),\nonumber \\
 C_n=&~P_{\text{d}_n}(1-P_{\text{f}_n}) - P_{\text{f}_n}(P_{\text{d}_n}-P_{\text{f}_n}),\nonumber \\
 B_n =& ~P_{\text{d}_n}(1-P_{\text{d}_n}), ~~D_n = P_{\text{f}_n}(1-P_{\text{f}_n}).
 \end{align*}
\subsection{Error Probability Approximation}\label{approxx}
In this section, we provide  a closed-form approximate expression for $P_e$ in \eqref{pe1}. 
To find the approximate expression for $P_e$, we approximate $\Delta$ in \eqref{lrt1} using a similar Gaussian distribution approximation as we conducted in Section \ref{j_drive}.

In Section \ref{j_drive} we approximated the conditional pdf $f(y_n|h)$ with $f^G(y_n|h)={\cal N}(m_{n,h},\Upsilon_{n,h}^2)$, where the mean $m_{n,h}$ and the variance $\Upsilon^2_{n,h}$ of the approximate distribution are provided in \eqref{G_par}. Relying on this Gaussian distribution approximation, we can also approximate the conditional pdf $f(\boldsymbol{y}|h)$. In particular, since the received signals at the FC, conditioned on $h$, are independent across sensors (see (\ref{y/H}-a)), we can approximate $f(\boldsymbol{y}|h)$ with  $f^G(\boldsymbol{y}|h)={\cal N}(\varphi_h,\Lambda_h)$, where $\varphi_h$ and $\Lambda_h$ are the mean vector and the {\it diagonal} covariance matrix with elements $m_{n,h}$ and $\Upsilon_{n,h}^2$, respectively.
Using this Gaussian distribution approximation, we can approximate $\Delta$ in \eqref{lrt1} as
\begin{align}\label{clt_lrt}
 \Delta\approx&\log \left( \frac{f^G(\boldsymbol{y}|h=1)}{f^G(\boldsymbol{y}|h=0)}\right)\\=&\log  \left(\frac{\sqrt{\det\Lambda_0} \text{exp}\left(-\frac{1}{2}(\boldsymbol{y}-\varphi_1)^T\Lambda_1^{-1}(\boldsymbol{y}-\varphi_1)\right)}{\sqrt{\det\Lambda_1}\text{exp}\left(-\frac{1}{2}(\boldsymbol{y}-\varphi_0)^T\Lambda_0^{-1}(\boldsymbol{y}-\varphi_0)\right)}\right)\nonumber\\=& R \!-\!\frac{1}{2}(\boldsymbol{y}\!-\!\varphi_1)^T\Lambda_1^{-1}(\boldsymbol{y}\!-\!\varphi_1) + \frac{1}{2}(\boldsymbol{y}\!-\!\varphi_0)^T\Lambda_0^{-1}(\boldsymbol{y}\!-\!\varphi_0),\nonumber
\end{align}
where $R=\log \left(\frac{\sqrt{\det\Lambda_0}}{\sqrt{\det\Lambda_1}}\right)$. Since the covariance matrices $\Lambda_0$ and $\Lambda_1$ are diagonal, the approximate expression for $\Delta$ in \eqref{clt_lrt} can be rewritten as 
\begin{align}
\label{delta1}
   & \Delta\approx R+  \frac{1}{2}\Delta'_N, ~\Delta'_N = \sum_{n=1}^N z_n, \nonumber\\
   &z_n = \frac{(y_n-m_{n,0})^2}{\Upsilon_{n,0}^2} - \frac{(y_n-m_{n,1})^2}{\Upsilon_{n,1}^2},
\end{align}
With the Gaussian distribution approximation, the optimal fusion rule in \eqref{u0_lrt} can be approximated with
\begin{equation}\label{del_N}
u_0= \begin{cases}
1, ~~\Delta'_N > \tau',\\
0, ~~\Delta'_N < \tau', 
\end{cases}
\end{equation}
where $\Delta'_N$ is given in \eqref{delta1} and $\tau'= 2(\tau-R)$. The error probability corresponding to the fusion rule in \eqref{del_N} is
\begin{equation}\label{Pe-Delta'}
P_e=\Pi_0 \Pr(\Delta'_N > \tau' |h=0)+\Pi_1 \Pr(\Delta'_N < \tau' |h=1).
\end{equation}
To find $P_e$ in \eqref{Pe-Delta'} we need the pdf of $\Delta'_N$ given $h$. We note that $z_n$ in \eqref{delta1} can be rewritten as a quadratic function of $y_n$
\begin{align}
\label{z_n2}
    z_n &=  a y_n^2 + b y_n +c, ~~\mbox{where}\nonumber\\
    a &= \frac{1}{\Upsilon^2_{n,0}}\!\!-\!\frac{1}{\Upsilon^2_{n,1}},~b= \frac{2m_{n,1}}{\Upsilon^2_{n,1}}\!-\!\frac{2m_{n,0}}{\Upsilon^2_{n,0}},~ c=\frac{m_{n,0}^2}{\Upsilon^2_{n,0}}\!-\!\frac{m_{n,1}^2}{\Upsilon^2_{n,1}}.
\end{align}
\begin{figure*}
 \begin{align}
\label{pdf_z}
    f(z_n|h) &= \frac{1}{g(z_n)}\left\{f^G_{y_n|h}\left(\frac{\Upsilon^2_{n,0}\Upsilon^2_{n,1}}{2}g(z_n)\!+\!m_{n,0}\Upsilon^2_{n,1}\!-\!m_{n,1}\Upsilon^2_{n,0}\right)\!+\!f^G_{y_n|h}\left(\frac{\!-\!\Upsilon^2_{n,0}\Upsilon^2_{n,1}}{2}g(z_n)\!+\!m_{n,0}\Upsilon^2_{n,1}\!-\!m_{n,1}\Upsilon^2_{n,0}\right)\right\},\nonumber\\
    g(z_n) &= \frac{2}{\Upsilon_{n,1}\Upsilon_{n,1}}\sqrt{(m_{n,0}-m_{n,1})^2+z_n(\Upsilon^2_{n,1}-\Upsilon^2_{n,0})}.
 \end{align}
 \hrulefill
\end{figure*}
Let $\mu_{z_n|h}$ and $\sigma^2_{z_n|h}$, denote the mean and variance of $z_n$ in \eqref{z_n2} given $h$, respectively. 
To find $\mu_{z_n|h}, \sigma^2_{z_n,h}$ we recall the following fact.\par \textbf{Fact}: Let $x \sim N (\mu, \sigma^2)$ be a Gaussian random variable with the mean $\mathbb{E}\{x\}=\mu$ and the variance $\sigma^2=\mathbb{E}\{x^2\}-\mu^2$. Then we have \cite{xx}:
\begin{align}
    \mathbb{E}\{x^2\} &= \mu^2+\sigma^2,\\
    \mathbb{E}\{x^3\} &=\mu(\mu^2+3\sigma^2),\nonumber\\
    \mathbb{E}\{x^4\}&=\mu^4+6\mu^2\sigma^2+3\sigma^4.\nonumber
\end{align}
Using this fact, we find
\begin{align}
&\mu_{z_n|h}=a(m_{n,h}^2+\Upsilon^2_{n,h})+b\,m_{n,h}+c, \\
&\sigma^2_{z_n|h}= 2 a^2( 2 m_{n,h}^2+\Upsilon^4_{n,h})+b \Upsilon^2_{n,h}(b+4\,a\,m_{n,h}),\nonumber
\end{align}
where $a,b,c$ are given in \eqref{z_n2} and $m_{n,h}, \Upsilon^2_{n,h}$ are given in \eqref{G_par}.
Relying on the Gaussian distribution approximation of $y_n$ given $h$, we can derive the pdf of $z_n$ given $h$, where the pdf expression is provided in  \eqref{pdf_z}. Since given $h$, $z_n$'s are independent, the pdf of $\Delta'_N$ given $h$, is convolution of these $N$ individual pdfs, which does not have a closed-form expression. This indicates that, even with the Gaussian distribution approximation, finding a closed-form expression of $P_e$ in \eqref{Pe-Delta'} for finite $N$ remains {\it elusive}. Hence, we resort to the {\it asymptotic regime} when $N$ grows very large and invoke the central limit theorem (CLT) to approximate $P_e$ in \eqref{Pe-Delta'}.
\par Lindeberg CLT is a variant of CLT, where the random variables are independent, but not necessarily identically distributed \cite{lin_1}. Let $\mu_{\Delta'_N|h}$ and $\sigma^2_{\Delta'_N|h}$ indicate the mean and variance of $\Delta'_N$ in \eqref{delta1} given $h$. We have $\mu_{\Delta'_N|h} = \sum_{n=1}^N  \mu_{z_n|h} $ and $\sigma_{\Delta'_N|h}^2 = \sum_{n=1}^N \sigma^2_{z_n|h}$. Assuming Lindeberg's condition, given below, is satisfied 
\begin{equation}
\lim_{N\rightarrow\infty}\frac{1}{\sigma_{\Delta'_N|h}^2}\sum_{n=1}^{N}\mathbb{E}\{(z_n - \mu_{z_n|h})^2\}=0,
\end{equation}
then, as $N$ goes to infinity, the normalized sum $(1/\sigma_{\Delta'_N|h}^2 ) \sum_{n=1}^N(z_n - \mu_{z_n|h})$ converges in distribution toward the standard normal distribution
\begin{equation}\label{lindeberg-CLT}
\frac{1}{\sigma_{\Delta'_N|h}^2} \sum_{n=1}^N (z_n - \mu_{z_n|h}) \overset{d}\rightarrow \mathcal{N}(0,1),
\end{equation}
where $\overset{d}\rightarrow$ indicates convergence in distribution. Using \eqref{lindeberg-CLT} we can approximate $P_e$ in \eqref{Pe-Delta'} using $Q$-function
\begin{equation}
\label{clt_pe}
     P_e=\Pi_0Q\left(\frac{\tau'- \mu_{\Delta'_N|0}}{\sigma^2_{\Delta'_N|0}}\right)+\Pi_1\!\left[1\!-\!Q\left(\frac{\tau'-\mu_{\Delta'_N|1}}{\sigma^2_{\Delta'_N|1}}\right)\right].
\end{equation}
\section{Formulating Our  Optimization Problem}\label{cost_fun}
As we stated before, 
our objective is to find the scale factors 
$\{c_{n,l}\}_{l=0}^{L-1}$ 
and the quantization thresholds 
$\{\mu_{n,l}\}_{l=1}^{L-1}$
in the transmit power control strategy  (\ref{alpha})  for all sensors such that the total $J$-divergence at the FC is maximized, subject to an average transmit power per sensor constraint.
We formulate the optimization problem, via writing the cost function and the constraints in terms of the optimization variables.
Recall total $J$-divergence at the FC is $J_{tot}=\sum_{n=1}^N J_n$, where $J_n$ in given in \eqref{sim_j}, and transmit power per sensor ${\cal P}_n$ is given in \eqref{alpha}. We note that $J_n$ depends on $g_n$ value, whereas ${\cal P}_n$ depends on the quantization interval to which $g_n$ belongs. The dependency of $J_n$ on $g_n$ stems from the fact that the FC has full knowledge of all channel gains $g_n$'s, and the optimal Bayesian fusion rule utilizes this full information. Hence, the error probability $P_e$ and its bound $J_{tot}$ depend on this full information. On the other hand, sensor $n$ only knows the  quantization interval to which $g_n$ belongs, and adapts its transmit power ${\cal P}_n$ according to this {\it partial} knowledge as well as its battery state. 
We seek the best $\{c_{n,l}\}_{l=0}^{L-1}$ 
and $\{\mu_{n,l}\}_{l=1}^{L-1}$, such that the solutions we obtain that do not depend on the specific channel gain realizations. Hence, we take the average of $J_n$ and ${\cal P}_n$ over $g_n$, conditioned that $g_n \in [\mu_{n,i},\mu_{n,i+1})$. By taking such a conditional average over $g_n$, the solutions we obtain do not depend on the specific channel gain realizations and are valid, as long as the channel gain statistics remain unchanged. The problem can be solved {\it offline} and its solutions can become available a {\it a priori} at the FC and the sensors. 
Let $\bar{J}_n^{(i)} \!=\!\mathbb{E}\{ J_n | g_n \in [\mu_{n,i}, \mu_{n, i+1} )\}$ and $ \bar{{\cal P}}_n^{(i)}\!=\!\mathbb{E}\{{\cal P}_n | g_n \in [\mu_{n,i}, \mu_{n, i+1}) \}$, respectively, denote the expectations of $J_n$ and ${\cal P}_n$ over $g_n$ and, conditioned that $g_n \in [\mu_{n,i}, \mu_{n,i+1})$.
In the following, we compute the two conditional expectations $\bar{J}_n^{(i)}$ and $\bar{{\cal P}}_n^{(i)}$, in terms of the optimization variables. 
To compute $\bar{J}_n^{(i)}$ we use the following fact.
\par {\bf Fact}: Suppose random variable $x$ has an exponential distribution with parameter $\lambda$, i.e., the pdf of $x$ is $f(x)=\lambda e^{-\lambda x}$. Consider the function $h(x) = \frac{a + b x}{c + d x}$, with given constants $a$, $b$, $c$ and $d$. Then, the average of $h(x)$, conditioned on $x$ being in the interval $[\mu_i,\mu_{i+1})$ is
\begin{align}
  &\mathbb{E}\{h(x) | x \in [\mu_i,\mu_{i+1})\}=\int_{\mu_i}^{\mu_{i+1}} h(x)f(x)dx\nonumber\\ 
    &= \frac{1}{d}\Big[a\beta(\mu_{i+1})-\frac{bc}{d}\beta(\mu_{i+1})-be^{-\lambda \mu_{i+1}}\nonumber\\ 
    &~~~~~~~~~~~~~~~~-a\beta(\mu_{i})-\frac{bc}{d}\beta(\mu_{i})-be^{-\lambda \mu_{i}}\Big],\nonumber
\end{align}
where
 \begin{align}
     \beta(x) &=\lambda \text{exp}{(\frac{c\lambda}{d})}~ \text{Ei} \Big(-\lambda x - \frac{c\lambda}{d}\Big),~~
     \text{Ei}(z)&=\int^{\infty}_{-z} \frac{e^{-t}}{t} dt.\nonumber 
 \end{align}
Using this fact and letting $a_1=a_2=c_1=c_2=\sigma^2_{w_n}$, $b_1=A_n \alpha_n^2,~ b_2= C_n \alpha_n^2$, $d_1=B_n \alpha_n^2$ and $d_2=D_n \alpha_n^2$, where $A_n, B_n, C_n, D_n$ are given in \eqref{sim_j},  we reach at
\begin{align}\label{ave_J}
  \bar{J}_n^{(i)} = \sum_{k=0}^K\phi_{n,k}\pi_{n,i}\Big[\Omega(\lfloor c_{n,i} k\rfloor,\mu_{n,i+1}^2)-\Omega(\lfloor c_{n,i} k\rfloor,\mu_{n,i}^2)\Big],
\end{align}
where the two dimensional function $\Omega(x,y)$ in \eqref{ave_J} is
\begin{align}\label{omega-f}
 &\Omega(x,y)\triangleq
  \nonumber\\&\frac{1}{B_n x}\Big[\sigma^2_{w_n}\beta_1(x,y)-\frac{A_n}{B_n}\sigma^2_{w_n}\beta_1(x,y) - A_n x  e^{(-y\gamma_{g_n} )}\Big]+
  \nonumber\\&
  \frac{1}{D_n x}\Big[\sigma^2_{w_n}\beta_2(x,y) - \frac{C_n}{D_n}\sigma^2_{w_n}\beta_2(x,y) - C_n x e^{(-y\gamma_{g_n} )}\Big],
\end{align}
and the two dimensional functions $\beta_1(x,y)$ and $\beta_2(x,y)$  in \eqref{omega-f} are
\begin{align*}
&\beta_1(x,y)\triangleq\gamma_{g_n}\text{exp}\Big(\frac{\sigma^2_{w_n}\gamma_{g_n}}{x B_n} \Big)\text{Ei}\Big(-\gamma_{g_n}y-\frac{\sigma^2_{w_n}\gamma_{g_n}}{x B_n}\Big),\nonumber
\end{align*}
\begin{align*}
&\beta_2(x,y)\triangleq\gamma_{g_n}\text{exp}\Big(\frac{\sigma^2_{w_n}\gamma_{g_n}}{x D_n}\Big)\text{Ei}\Big(-\gamma_{g_n}y-\frac{\sigma^2_{w_n}\gamma_{g_n}}{x D_n}\Big). 
\end{align*}
We can compute $\bar{{\cal P}}_n^{(i)}$ using \eqref{alpha} as the following
\begin{equation}\label{ave_alph}
    \bar{{\cal P}}_n^{(i)}=\widehat{\Pi}_{n,1}\sum_{k=0}^{K}\phi_{n,k}\pi_{n,i}\lfloor c_{n,i} k\rfloor 
\end{equation}
We formulate our problem, denoted as (P1), as the following 
%
\leqnomode
\begin{align*}\tag{P1}
&\max_{\{c_{n,l}\}_{l=0}^{L-1},\{\mu_{n,l}\}_{l=1}^{L-1},{\forall n}}~ \sum_{n=1}^N\sum_{i=0}^{L-1}\bar{J}_n^{(i)}\nonumber\\
  {\hbox{s.t.}}~ & c_{n,l} \in [0,1], ~l=0, ..., L-1, \forall n\\
  & 0< \mu_{n,l}  < \infty, ~ l=1, ..., L-1, \forall n  \\
  & \sum_{i=0}^{L-1}\  \bar{{\cal P}}_n^{(i)}\leq {\cal P}_0,~\forall n \\
  & \boldsymbol{\Phi}_{n} = -(\boldsymbol{\Psi}_n^T-\textbf{I}-\textbf{B})^{-1}\textbf{1},~\forall n 
\end{align*}
\reqnomode

Regarding the implementation of (P1) a remark follows. 

{\bf Remark:} We note that cost function and the constraints in (P1) are {\it decoupled} across sensors. Hence, (P1) can be decomposed into $n$ sub-problems, denoted as (P2), as the following
%
\leqnomode
\begin{align*}\tag{P2}
&\max_{\{c_{n,l}\}_{l=0}^{L-1},\{\mu_{n,l}\}_{l=1}^{L-1}} ~\sum_{i=0}^{L-1}\bar{J}_n^{(i)}\nonumber\\
 {\hbox{s.t.}}~ & c_{n,l} \in [0,1], ~l=0, ..., L-1, \\
  & 0< \mu_{n,l}  < \infty, ~ l=1, ..., L-1,   \\
  & \sum_{i=0}^{L-1}\  \bar{{\cal P}}_n^{(i)}\leq {\cal P}_0,\\
  & \boldsymbol{\Phi}_{n} = -(\boldsymbol{\Psi}_n^T-\textbf{I}-\textbf{B})^{-1}\textbf{1}.
\end{align*}
%
%
%
This implies that solving (P1) is equivalent to solving (P2) $N$ times  for $n=1,...,N$. 
It also implies that solving (P1)
can lend itself to a {\it distributed} implementation, where sensor $n$ solves its corresponding (P2) independent of the other sensors. 
For implementing our proposed power control strategy, we assume that the FC solves (P1) once.
Based on the obtained solution, in each time slot $t$ the FC quantizes $g_{n,t}$'s and informs sensor $n$ of the quantization interval to which $g_{n,t}$ belongs, via a limited feedback channel. Sensor $n$ solves its corresponding
(P2) once, and based on the obtained solution it
sets its transmit power control strategy in \eqref{alpha} once. Then, in each time slot $t$ sensor $n$
chooses its transmit power ${\cal P}_{n,t}$ according to \eqref{alpha},
considering its battery state and the received feedback information. 
It is worth mentioning the  difference between optimizing the total $J$-divergence and the approximate $P_e$ expression in \eqref{clt_pe}. 
%
%
Different from (P1),  the approximate $P_e$ expression in \eqref{clt_pe} {\it cannot be decoupled} across sensors. Therefore, constrained minimization of $P_e$ does not render itself to a distributed implementation, i.e. each sensor needs to solve (P1), with $P_e$ in \eqref{clt_pe} being the cost function, which ensues a much higher computational complexity.   
\section{Solving Problem (P1)} \label{how-to-solve-P1}
%
Since solving (P1) is equivalent to solving (P2) $N$ times, in this section we focus on solving (P2). 
Let examine how the cost function and the constraints in (P2) depend on the optimization variables. 

$\bullet$ {Dependency of $\bar{J}_n^{(i)}$}: Considering \eqref{ave_J}, its explicit dependency on $\{c_{n,l}, \mu_{n,l}\}$'s is clear. It also depends implicitly on $\{c_{n,l}, \mu_{n,l}\}$'s through the probabilities $\pi_{n,l}$'s and the vector entries $\phi_{n,k}$'s. Recall that $\phi_{n,k}$'s are the entries of vector $\boldsymbol{\Phi}_{n}$ given in (\ref{inverse_battery}). This vector depends on the matrix $\boldsymbol{\Psi}_{n}$, whose entries are given in (\ref{pij}) and depend on $\{c_{n,l}, \mu_{n,l}\}$'s. 

$\bullet$ {Dependency of $\bar{{\cal P}}_n^{(i)}$}: Considering \eqref{ave_alph}, its explicit dependency on $c_{n,l}$'s is clear. It also depends implicitly on $\{c_{n,l}, \mu_{n,l}\}$'s through  $\pi_{n,l}$'s and $\phi_{n,k}$'s.

$\bullet$ {Dependency of $\boldsymbol{\Phi}_{n}$}: It depends implicitly on $\{c_{n,l}, \mu_{n,l}\}$'s. 
We note that problem (P2) is {\it not concave} with respect to the optimization variables. Moreover, the objective function and the constraints in
(P2) are {\it not differentiable} with respect to the optimization variables. Hence,
existing gradient-based algorithms for solving non-convex optimization problems cannot be used to solve (P2).
\subsection{Deterministic Search Method}\label{deterministic-search}
We resort to a grid-based search method, which requires $(2L-1)$-dimensional
search over the search (parameter) space $[0,1]^{L}\times(0,\infty)^{L-1}$. 
To curb the
computational complexity of this grid-based search, we can limit $\mu_{n,l}$'s  to a maximum value, denoted as $\mu_{max}$. We refer to the
solution obtained from solving (P2)  using this method the {\it optimal} solution, in the sense that it is the {\it best} attainable solution for (P2). Clearly, the accuracy of this solution depends on the resolution of the grid-based search. Suppose the intervals $[0,1]$ and $(0,\mu_{\max}]$ are divided into $N_c$ and $N_{\mu}$ sub-intervals, respectively.
Therefore, the search space of (P2), denoted as ${\cal D}$,  consists of $(N_c)^{L} (N_{\mu})^{L-1}$ discrete points in the original $(2L-1)$-dimensional search space. 
To find the computational complexity of obtaining the {\it optimal} solution for (P2), we note that the solver unit (either FC or sensor $n$) 
needs to perform two tasks for each point in ${\cal D}$: task (i)  forming $\boldsymbol{\Psi}_n$ and solving \eqref{inverse_battery} to find $\boldsymbol{\Phi}_n$, task (ii) calculating $\bar{J}_n^{(i)}$ and $\bar{{\cal P}}_n^{(i)}$. Our numerical results show that for a fixed $\{c_{n,l}\}_{l=0}^{L-1}$ and $\{\mu_{n,l}\}_{l=1}^{L-1}$ the computational complexity of task (i) and task (ii) are $\mathcal{O}(K^{3.2})$ and  $\mathcal{O}(K^{1.1})$, respectively.  %
Hence, the computational complexity of finding the {\it optimal} solution for (P2) is   $\mathcal{O}\left( N_c^{L} N_{\mu}^{L-1}(K^{3.2}+K^{1.1})\right)$. Since $K^{1.1}$ order is dominated by $K^{3.2}$, the computational complexity of finding the {\it optimal} solution for (P2)  can be simplified to $\mathcal{O}\left( N_c^{L} N_{\mu}^{L-1}K^{3.2}\right)$.   
\subsection{Random Search Method}\label{random-method}
%
%
Finding the {\it optimal} solution of (P2) using the  grid-based search, as described above, requires searching search space ${\cal D}$ {\it deterministically}. 
In contrast, in a {\it random} search algorithm, only a randomly chosen subset of the points in ${\cal D}$ is searched to find a solution. The size of this subset can be chosen to be  smaller than  $(N_c)^{L} (N_{\mu})^{L-1}$, and hence, the computational complexity of finding a solution using a random search algorithm can be significantly lowered. 
We refer to the solution obtained from solving (P2) using a random search algorithm the {\it c-optimal} solution, in the sense that it is a close-to-optimal solution. 

Among the random search algorithms in the literature, we choose the so-called 
  ``Recursive Random Search (RRS) algorithm" \cite{recursive}.
%
Our reason for this choice is that the authors in  \cite{recursive} showed that RRS algorithm outperforms significantly the traditional search algorithms (e.g., genetic algorithms, multi-start hill climbing algorithms, and simulated annealing algorithm) for most optimization problems. RRS algorithm  consists of two phases: exploration (global) phase and exploitation (local) phase. In exploration phase, the algorithm performs random sampling from the entire sample space ${\cal D}$, to inspect the overall form of the objective function, and to 
identify  ``promising areas''  in ${\cal D}$ \cite{recursive}. 
In exploitation phase, the algorithm continues to search {\it only} within the identified ``promising areas'', using recursive  random sampling. As the search continues, the sample space is shrunk gradually (according to
the previously drawn samples),  and the 
algorithm learns more details of the objective function,
until it finally converges to a local optimum,  which will be considered as the solution of the optimization problem in hand \cite{recursive}.
For our work to be self-contained, in the following we overview RRS algorithm, with reference to the lines in the pseudo-code of Algorithm \ref{Alg1}.   

$\bullet$ {\it Exploration Phase}: 
To describe this phase and to illustrate the efficiency of RRS algorithm in finding the solution of (P2), we need to first introduce the following notations and concepts. Suppose $x \!=\! [c_{n,0},...,c_{n,L-1},\mu_{n,1},...\mu_{n,L-1}]$ denote a sample (point) in  $\mathcal{D}$, and $ J_{min}, J_{max}$ indicate the minimum and the maximum values of the objective function, respectively. 
We define the distribution function of the objective function values as 
$r=\frac{m(A_{\mathcal{D}}(r))}{m(\mathcal{D})}$, for $r \! \in \![0,1]$, 
where $m(.)$ denotes the cardinality of the set.  Given $r$ value, set  $A_{\mathcal{D}}(r) \subset \mathcal{D}$ with the cardinality  $m(A_{\mathcal{D}}(r))= r \times  m(\mathcal{D})$ is the set of points in $\mathcal{D}$  whose values of the objective function exceed a threshold $ J_{tr} \in [J_{min}, J_{max}]$.
\begin{equation*}
    A_{\mathcal{D}}(r)\! =\! \left\{\!x\! \in\! \mathcal{D}|\sum_{i=0}^{L-1}\bar{J}_n^{(i)}(x)\! \geq\! J_{tr}(r)\!\right\}\!, m(A_{\mathcal{D}}(r))\!=\! r\! \times\!  m(\mathcal{D})
\end{equation*}
For this reason $A_{\mathcal{D}}(r)$ is called the \textit{r-percentile} set in  $\mathcal{D}$ \cite{recursive}. We note that $A_{\mathcal{D}}(1)=\mathcal{D}$ and $\lim_{r \rightarrow 0} A_{\mathcal{D}}(r)$ converges to the global optimum of the problem \cite{recursive}.  
Now, consider the \textit{r-percentile} set $A_{\mathcal{D}}(r)$ in  $\mathcal{D}$ and its corresponding $J_{tr}(r)$ value. The goal in exploration phase is to reach a point in $A_{\mathcal{D}}(r)$ with probability $p$, via random sampling. 
%
The question is: how many random samples of $\mathcal{D}$ should we draw, such that we reach a point in $A_{\mathcal{D}}(r)$ with probability $p$?

To answer this question, let $\boldsymbol{X}\!=
\!\{x_j\}_{j=1}^{Q_1}$ be the set of randomly drawn samples from ${\cal D}$ 
%
that satisfy the average transmit power constraint in (P2), 
%
and $x_j^* \! \in \! \boldsymbol{X}$  provides the largest value of the objective function. We have %
%
%
\begin{equation*}
    \label{p-r-Q}
    p=\Pr\left(x_j^* \in A_{\mathcal{D}}(r)\right)=1\!-\!\Pr\left(x_j^*\notin A_{\mathcal{D}}(r)\right)\!=\! 1\!-\!(1\!-\!r)^{Q_1}.
\end{equation*}
Solving $p$ for $r$ we reach at $r=1- (1-p)^{1/Q_1}$. Solving $p$ for $Q_1$  we obtain  $Q_1=\frac{\ln (1-p)}{\ln (1-r)}$.  For any probability value $p$, as  $Q_1$ increases, $r$ tends to $0$ and $\lim_{r \rightarrow 0} A_{\mathcal{D}}(r)$ converges to the global maximum of (P2).  
%
%

Lines 2,3,4 of the pseudo-code correspond to this phase. 
%
%
We take $Q_1$ random samples from ${\cal D}$, each denoted as $x_{q_1}$, and 
put them in $\boldsymbol{X}_t\!=
\!\{x_{q_1}\}_{{q_1}=1}^{Q_1}$ and initialize $\boldsymbol{X}\!=\{\}$. For each sample $x_{q_1} \in \boldsymbol{X}_t$, we check whether  the average transmit power constraint is held. If the constraint is satisfied, $x_{q_1}$ is added to $\boldsymbol{X}$. If the constraint is not satisfied, 
 we take another sample from the set $\mathcal{D}\backslash \boldsymbol{X}_t$ and add this new sample to  $\boldsymbol{X}_t$. We repeat this procedure until
 %
 %
 $m(\boldsymbol{X})$ reaches  $Q_1$. Using the samples in $\boldsymbol{X}\!=\!\{x_j\}_{j=1}^{Q_1}$, the algorithm computes the threshold
$J_{tr}$.
%
Having the set $\boldsymbol{X}$, whose elements represent the ``promising areas'' in $\mathcal{D}$, the algorithm enters exploitation phase. Any future sample we encounter in the next phase that has a greater value of the objective function than  $J_{tr}$  belongs to $A_{\mathcal{D}}(r)$.


$\bullet$ {\it Exploitation Phase}:
 Consider $\boldsymbol{X}\!=\!\{x_j\}_{j=1}^{Q_1}$. For each sample $x_j \! \in \! \boldsymbol{X}$ we first determine several neighborhoods\footnote{  The neighborhood  $N_{\rho}(x_j)$, for $\rho=1,...,\rho_0$,  is the set of samples that are neighbors of $x_j$. Its size $m(N_{\rho}(x_j))$ depends on the dimensionality of search space ${\cal D}$ ($(2L-1)$ here) and the resolution of the grid (parameters $N_c, N_{\mu}$ here). To identify different neighborhoods of $x_j$, we have used MATLAB's  function {\it neighbourND}. 
 For instance, for $L\!=\!2$ and 
  $N_c\!=\! 10, N_{\mu}\!=\! 100$ we have  $m(N_1(x_j)) \!=\!17$, $m(N_2(x_j)) \!=\! 60$, $m(N_3(x_j)) \!=\!139$. 
  }
 %
 $N_{\rho}(x_j)$ for $\rho\!=\!1,...,\rho_0$, such that $m(N_{\rho}(x_j)) \! < \! m(N_{\rho+1}(x_j))$. Given the parameter\footnote{To enable efficient random search even in the smallest neighborhood we choose  $Q_2 < m(N_1(.))$.} $Q_2$, the description of the recursive random search in these neighborhoods to find the solution of (P2) follows. 

 For each sample $x_j \! \in \! \boldsymbol{X}$, 
we start by letting the search space be ${\cal S} \!=\! N_{\rho_0}(x_j)$, and search ${\cal S}$ hoping to to find a {\it better} sample than $x_j$. In particular, we take $Q_2$ random samples from $N_{\rho_0}(x_j)$, each denoted as $x_{q_2}$, and put them in $\boldsymbol{Y}_t \!=\!\{x_{q_2}\}_{q_2=1}^{Q_2}$ and initialize $\boldsymbol{Y}_j=\{\}$. For each sample $x_{q_2}\!\in \! \boldsymbol{Y}_t$, we check two conditions: (i) whether the average transmit power constraint is held, (ii) whether the objective function evaluated at $x_{q_2}$ provides a lager value than $J_{tr}$. If both constraints are satisfied, $x_{q_2}$ is added to $\boldsymbol{Y}_j$. After checking all samples in $\boldsymbol{Y}_t$ we
examine  $\boldsymbol{Y}_j$. Depending on whether $\boldsymbol{Y}_j \! \neq \! \{\}$, meaning there exists at {\it least one  better} sample than $x_j$ in ${\cal S}$, or
$\boldsymbol{Y}_j \! = \! \{\}$, meaning {\it no better} sample than $x_j$ is found in ${\cal S}$, we take two different actions. 

If $\boldsymbol{Y}_j \! \neq \! \{\}$ we select the sample in $\boldsymbol{Y}_j$ that provides the largest value of the objective function, denoted as $x_i^*$, and replace $x_j \! \in \! \boldsymbol{X}$ with $x_i^* $, and change ${\cal S}$ from   $N_{\rho_0}(x_j)$  to $N_{\rho_0}(x_i^*)$, and continue with searching  the new ${\cal S}$. This procedure of changing the center of ${\cal S}$ (without shrinking it) in exploitation phase is called ``re-align sub-phase'' \cite{recursive}. 
%
%
%
%
%
%
%
However, if $\boldsymbol{Y}_j \! = \! \{\}$ we shrink ${\cal S}$ by changing ${\cal S}$ to $N_{\rho_0-1}(x_j)$. 
This procedure of shrinking ${\cal S}$ (without changing its center) in exploitation phase is called ``shrink sub-phase'' \cite{recursive}. When searching $N_{\rho_0-1}(x_j)$, if we find a {\it better} sample than $x_j$, we replace $x_j \! \in \! \boldsymbol{X}$ with this {\it better} sample. Otherwise, we further shrink ${\cal S}$ by changing ${\cal S}$ to $N_{\rho_0-2}(x_j)$.
%
 We alternatively perform re-align and shrink sub-phases for $x_j$, until we get to search the smallest neighborhood
$N_1(.)$ of a sample. 
Note that we limit the number of times we perform re-align sub-phase during the exploitation procedure for $x_j$ to $Q_1$, relying on the fact that after drawing $Q_1$ samples from ${\cal D}$ we reach a point in  $A_{\mathcal{D}}(r)$ with probability $p$. 
%
%
At this point, the exploitation procedure for $x_j$ ends, and $x_j \in \boldsymbol{X}$ is either {\it kept unchanged} or {\it replaced with a better sample} that is found during its exploitation procedure. We repeat the exploitation procedure for all samples in $\boldsymbol{X}$, and at the end we obtain a refined and fully exploited $\boldsymbol{X}$. We let the solution of (P2) be $\arg\max_{x_j \in \boldsymbol{X}}(\sum_{i=0}^L\bar{J}_n^{(i)}(x_j))$. %



%


To find the
computational complexity of obtaining the {\it c-optimal} solution for (P2), we note that during the exploitation phase the solver unit %
 needs to perform {\it repeatedly} the same two tasks, task (i) and task (ii) in Section \ref{deterministic-search}, with computational complexity $\mathcal{O}\left( K^{3.2}\right)$ and $\mathcal{O}\left( K^{1.1}\right)$, respectively. To find out the number of  repetition of tasks, we focus on the exploitation procedure for $x_j \!\in \! \boldsymbol{X}$. After  each performance of re-aligning ${\cal S}$ or shrinking  ${\cal S}$, we randomly search ${\cal S}$. i.e., we evaluate the objective function $Q_2$ times.
 Hence, the number of repetition of tasks for each $x_j \!\in \! \boldsymbol{X}$ is equal to $Q_2 \! \times \! \# (\mbox{performing re-align}) \! \times \! \# (\mbox{performing shrink})$. Since $\# (\mbox{performing re-align}) \! \leq \! Q_1$ and $\# (\mbox{performing shrink}) \! \leq \! \rho_0$,  the computational complexity corresponding to the exploitation procedure for $x_j \!\in \! \boldsymbol{X}$ is upper bounded by  $\mathcal{O}\left(  Q_2 Q_1 \rho_0   (K^{3.2}+K^{1.1})\right)$. Therefore, the computational complexity of finding the {\it c-optimal} solution for (P2) is upper bounded by
 $\mathcal{O}\left(  Q_2 Q_1^2 \rho_0   (K^{3.2}+K^{1.1})\right)$, which can be simplified to $\mathcal{O}\left(  Q_2 Q_1^2 \rho_0   K^{3.2}\right)$.

\begin{algorithm}[t!]
{\small
\caption{pseudo-code of RSS algorithm }\label{Alg1}
1: Initialization phase:
\begin{itemize}
    \item Set parameter space ${\cal D}$ with  $(N_c)^{L} \times(N_{\mu})^{L-1}$ points;
    \item Initialize exploration parameters $(p,r)$ and let $Q_1 = \ln{(1-p)}/\ln{(1-r)}$;
    \item Initialize exploitation parameter  $Q_2$ based on ($N_c, N_{\mu},L$);
\end{itemize}
2: Start exploration phase, take $Q_1$ uniform random $~~~$samples from ${\cal D}$ and put them in $\boldsymbol{X}_t=
\{x_{q_1}\}_{q_1=1}^{Q_1}$, $~~~$initialize $\boldsymbol{X}\!=\{\}$\; 
3: 
\Repeat
{$m(\boldsymbol{X})=Q_1$}
{\For{$x_{q_1}\in \boldsymbol{X}_t$}
{
\eIf{$\sum_{i=0}^{L-1} \bar{{\cal P}}_n^{(i)}(x_{q_1})\leq {\cal P}_0$}
{Put $x_{q_1}$ in $\boldsymbol{X}$\;}{Take another sample from $\mathcal{D}\backslash \boldsymbol{X}_t$ and add it to $\boldsymbol{X}_t$\;}
}}
4: Calculate the threshold using the samples in $~~~~\boldsymbol{X}=
\{x_j\}_{j=1}^{Q_1}$,  $J_{tr} = 1/Q_1\sum_{j=1}^{Q_1}\left(\sum_{i=0}^{L-1}\bar{J}_n^{(i)}(x_j)\right)$\;
5: Start exploitation phase,
determine the neighborhoods of $~~~~$sample $x_j$ as $N_1(x_j), N_2(x_j), ..., N_{\rho_0}(x_j)$\;
%
\For{$x_j\in \boldsymbol{X}$}
{Initialize $\boldsymbol{Y}_j=\{\}$, { $I=0$}, Take $Q_2$ uniform random samples from  $N_{\rho}(x_j)$ and put them in $\boldsymbol{Y}_t \!=\!\{x_{q_2}\}_{q_2=1}^{Q_2}$\;
\For{$x_{q_2}\in \boldsymbol{Y}_t $}{
\If{{\footnotesize $\sum_{i=0}^{L-1}\!\bar{J}_n^{(i)}(x_{q_2})\!\geq\!J_{tr}$\! \& \!$\sum_{i=0}^{L-1}\! \bar{{\cal P}}_n^{(i)}(x_{q_2})\!\leq\! {\cal P}_0$} }
{Add $x_{q_2}$ to $\boldsymbol{Y}_j$\;}}
\eIf{$\boldsymbol{Y}_j\neq\{\}$ { \& $I<Q_1$ }}
{  $x_i^*\!=\!\arg\max_{x_i \in \boldsymbol{Y}_j}(\sum_{i=0}^{L-1}\bar{J}_n^{(i)}(x_i))$,
  replace $x_j \in \boldsymbol{X}$ with $x_i^*$,
change the search space from $N_{\rho}(x_j)$  to $N_{\rho}(x_i^*)$\;
{ $I = I+1$\;}
}
{
change the search space  from $N_{\rho}(x_j)$  to $N_{\rho-1}(x_j)$ \;
%
%
} 
}
%
%
%
%
6: $x_{opt}\!=\!\arg\max_{x_j \in \boldsymbol{X}}(\sum_{i=0}^{L-1}\bar{J}_n^{(i)}(x_j))$\; 
}
\end{algorithm} 
\subsection{Hybrid Deterministic-Random Search Method}\label{heuristic}
In this section we propose a {\it hybrid} method to find the optimization variable $\{c_{n,l}, \mu_{n,l}\}$'s. In particular, we first obtain the quantization thresholds $\{\mu_{n,l}\}$'s using a different objective function. Then given the optimized $\{\mu_{n,l}\}$'s, we solve (P3), given below, using RSS algorithm.  
%
%
\leqnomode
\begin{align*}\tag{P3}
&\max_{\{c_{n,l}\}_{l=0}^{L-1}} ~\sum_{i=0}^{L-1}\bar{J}_n^{(i)}\nonumber\\
 {\hbox{s.t.}}~ & c_{n,l} \in [0,1], ~l=0, ..., L-1, \\
  & \sum_{i=0}^L  \bar{{\cal P}}_n^{(i)}\leq {\cal P}_0,\\
  &\boldsymbol{\Phi}_{n} = -(\boldsymbol{\Psi}_n^T-\textbf{I}-\textbf{B})^{-1}\textbf{1}.
\end{align*}
%
We refer to the solution we obtain using this hybrid method the {\it sub-optimal} solution, in the sense that it is worse than the optimal solution. The  {\it sub-optimal} solution is also worse than {\it c-optimal} solution for two reasons: (i) we detangle optimizing $\{\mu_{n,l}\}$'s and $\{c_{n,l}\}$'s, (ii) we use a different objective function to optimize $\{\mu_{n,l}\}$'s. 
The main advantage of using this hybrid method is that finding the 
{\it sub-optimal} solution has a lower computational complexity than that of  the
{\it c-optimal} solution. %
Our numerical results in Section \ref{simulation} show that the objective function values  at the {\it c-optimal} and the {\it sub-optimal} solutions are very close to each other and also very close to that of the {\it optimal} solution. 
%
%
%
%
%
In the following, we consider two different objective functions that we use to obtain the optimal $\{\mu_{n,l}\}$'s. To motivate these objective functions, we consider the input-output relationship of the quantizer in Section \ref{battery model} $\bar{g}_n=Q(g_{n})$. If the quantizer input  $g_{n}$ lies in the interval  $\mathcal{I}_{n,l}$ then the quantizer output is $\bar{g}_n=\mu_{n,l}$. The quantization error is $e_n=g_n - \bar{g}_n$. 
\subsubsection{Finding $\{\mu_{n,l}\}$'s via Minimizing Mean Absolute Error (MMAE)}\label{MAE}
%
%
The first objective function we consider is mean of absolute quantization error (MAE), denoted as $\mathbb{E} \{|g_n-\bar{g}_n|\}$. We can express MAE as follows.   
\begin{equation}
 \mathbb{E} \{|g_n-\bar{g}_n|\} = \sum_{l=0}^{L-1} \int^{\mu_{n,l+1} }_{\mu_{n,l}} (x-\mu_{n,l}) f_{g_n}(x)dx
\end{equation}
%
To find $\{\mu_{n,l}\}$'s that minimize MAE, %
we take the first derivative of MAE with respect to $\mu_{n,l}$ and set the derivative equal to zero.  We reach at 
\begin{equation}\label{F_mu}
  F_{g_n}(\mu_{n,l+1})= F_{g_n}(\mu_{n,l}) + (\mu_{n,l}-\mu_{n,l-1})f_{g_n}(\mu_{n,l}) 
\end{equation}
Recall $\mu_{n,0}\!=\!0$ and  $\mu_{n,L}\!=\!{ \infty}$,  and hence $F_{g_n}(0)\!=\!0$ and $F_{g_n}(\infty)=1$. 
We initiate $\mu_{n,1}$ and find $\mu_{n,2}$ using \eqref{F_mu}. Having $\mu_{n,1},\mu_{n,2}$, we find $\mu_{n,3}$ using \eqref{F_mu}. We repeat this until we find all $\{\mu_{n,l}\}$'s. At this point, we check whether the condition $F_{g_n}(\infty)=1$ is met. 
 If $F_{g_n}(\infty)$ is less (greater) than one, we increase (decrease) the initial value of $\mu_{n,1}$ and find a new set of values for $\{\mu_{n,l}\}$'s.
We continue changing the initial value of $\mu_{n,1}$ and finding new values for $\{\mu_{n,l}\}$'s, until the condition  $F_{g_n}(\infty)=1$ is satisfied. 
%
%
\subsubsection{Finding $\{\mu_{n,l}\}$'s via Maximizing output Entropy (MOE)}\label{MOE}
The second objective function we consider is the mutual information between $g_n$ and $\bar{g}_n$, denoted as $I(g_n; \bar{g}_n)$. We have 
 $I(g_n;\bar{g}_n) \!=\!  H(\bar{g}_n) \!-\!  H (\bar{g}_n|g_n)$,
where $H(x)$ denotes the entropy of discrete random variable
$x$. To find $\{\mu_{n,l}\}$'s that maximize $I(g_n; \bar{g}_n)$, we note that $H (\bar{g}_n|g_n)$ is zero, since $\bar{g}_n= Q(g_n)$ and hence, given $g_n$, $\bar{g}_n$ is also known.  Furthermore, $H(\bar{g}_n)$ is maximized when $\bar{g}_n$ follows a uniform distribution, i.e., we set $\pi_{n,l}\!=\!\Pr(\mu_{n,l} \! \leq \! g_n \!<\! \mu_{n,l+1}) \!=\! \frac{1}{L+1}$.
and the threshold $\mu_{n,l}$ can be obtained as $\mu_{n,l} \!=\!  \gamma_{g_n}\ln{\left(1-\frac{l}{L+1}\right)}$. 

The computational complexity of finding the {\it sub-optimal} solution for (P2) is the sum of two terms. The first term is the computational complexity of finding $\{c_{n,l}\}$'s using RRS algorithm in Section  \ref{random-method}, and is upper bounded by
 $\mathcal{O}\left(  Q_2 Q_1^2 \rho_0   (K^{3.2}+K^{1.1})\right)$. 
We note that   $Q_2$ in this section is chosen according to $m(N_1(.))$, which  depends on $(L, N_c)$, whereas $Q_2$ in Section \ref{random-method} is chosen according to $m(N_1(.))$, which  depends\footnote{ For instance, for $L\!=\!2$ and 
  $N_c\!=\! 10, N_{\mu}\!=\! 100$, we choose $Q_2 < 17$ in  Section \ref{random-method} and  we choose $Q_2 < 5$ here.} on $(L, N_c, N_{\mu})$. Hence, $Q_2$ here is smaller than $Q_2$ in Section \ref{random-method}. The second term is  the computational complexity of finding $\{\mu_{n,l}\}$'s optimizing one of the two objective functions in this section. 
The computational complexity of finding $\{\mu_{n,l}\}$'s via MMAE is negligible, due to the simplicity of solving \eqref{F_mu}. Our simulations show that for different $L$ values, solving \eqref{F_mu} takes only several msec. 
The computational complexity of finding $\{\mu_{n,l}\}$'s via MOE is almost zero, due to the available closed-form solutions. 
\section{Simulation results and discussion}\label{simulation}
We corroborate our analysis with MATLAB simulations and investigate: (i) the effect of the optimization variables on the objective function and the entries of $\boldsymbol{\Phi}$ in \eqref{inf},
%
(ii)  the accuracy of different search methods in
Section \ref{how-to-solve-P1} in solving (P2) as well as the existing trade-off between detection performance and average transmit power, 
%
%
(iii) the behavior of the optimized scale factors $\{c_l\}$'s with respect to the fading channel gain $g_n$, 
%
(iv) the accuracy of the $P_e$ approximate in  \eqref{clt_pe}.
%
%
(v) the dependency of the system error probability $P_e$ (achieved with the optimized variables) on  $K,\rho,L$, and the SNR corresponding to observation channel defined as SNR$_s\!=\!20\log(\mathcal{A}/\sigma_v)$.

$\bullet$ {\bf Effect of  optimization variables}:
\begin{figure}[!t]
 \begin{subfigure}[t]{0.24\textwidth}
 \centering
  \includegraphics[width=47mm]{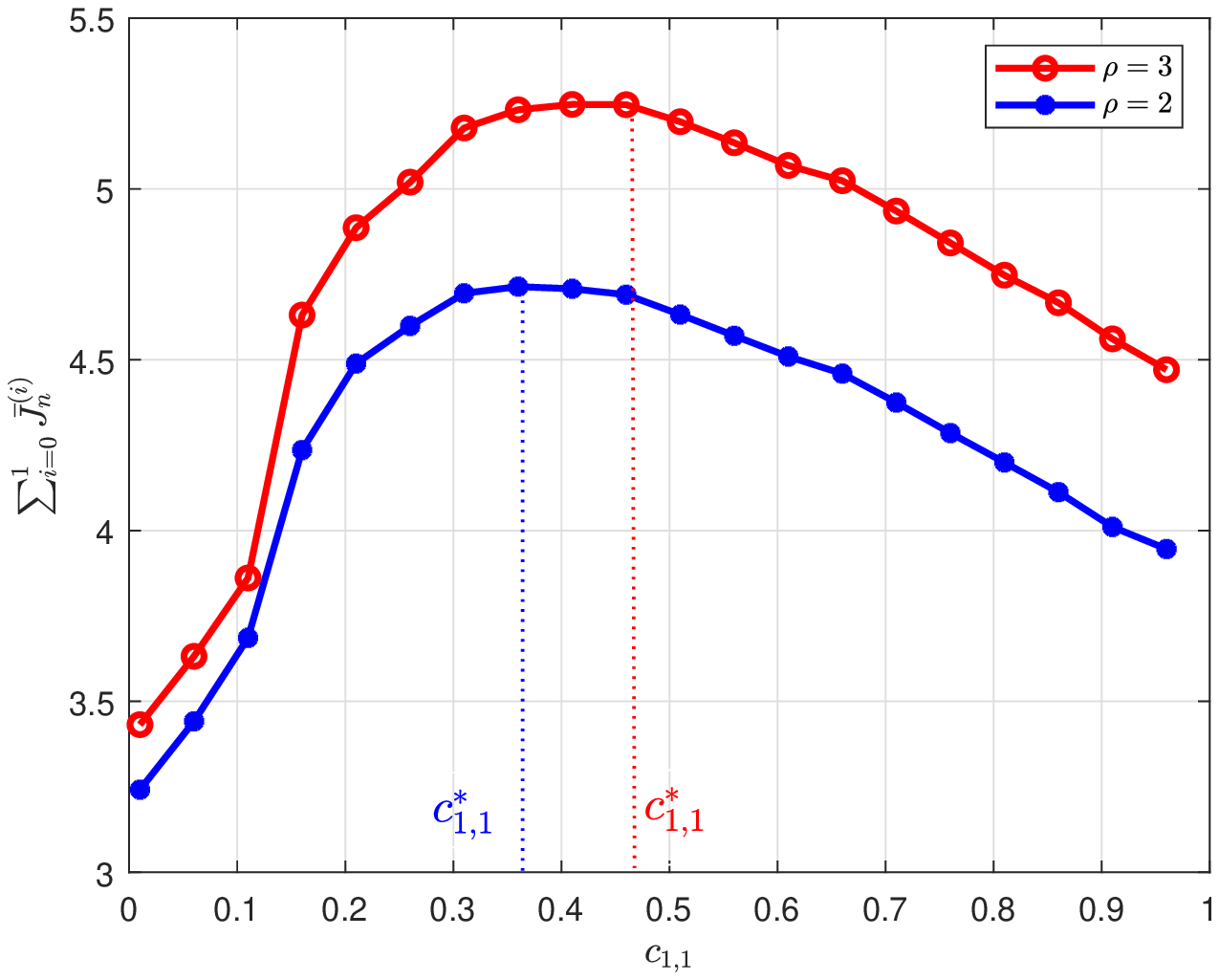}
 \caption{{$\mu_{1,1}=1.1$}}
\label{f6}
\end{subfigure}
\begin{subfigure}[t]{0.24\textwidth}
  \centering
  \includegraphics[width=47mm]{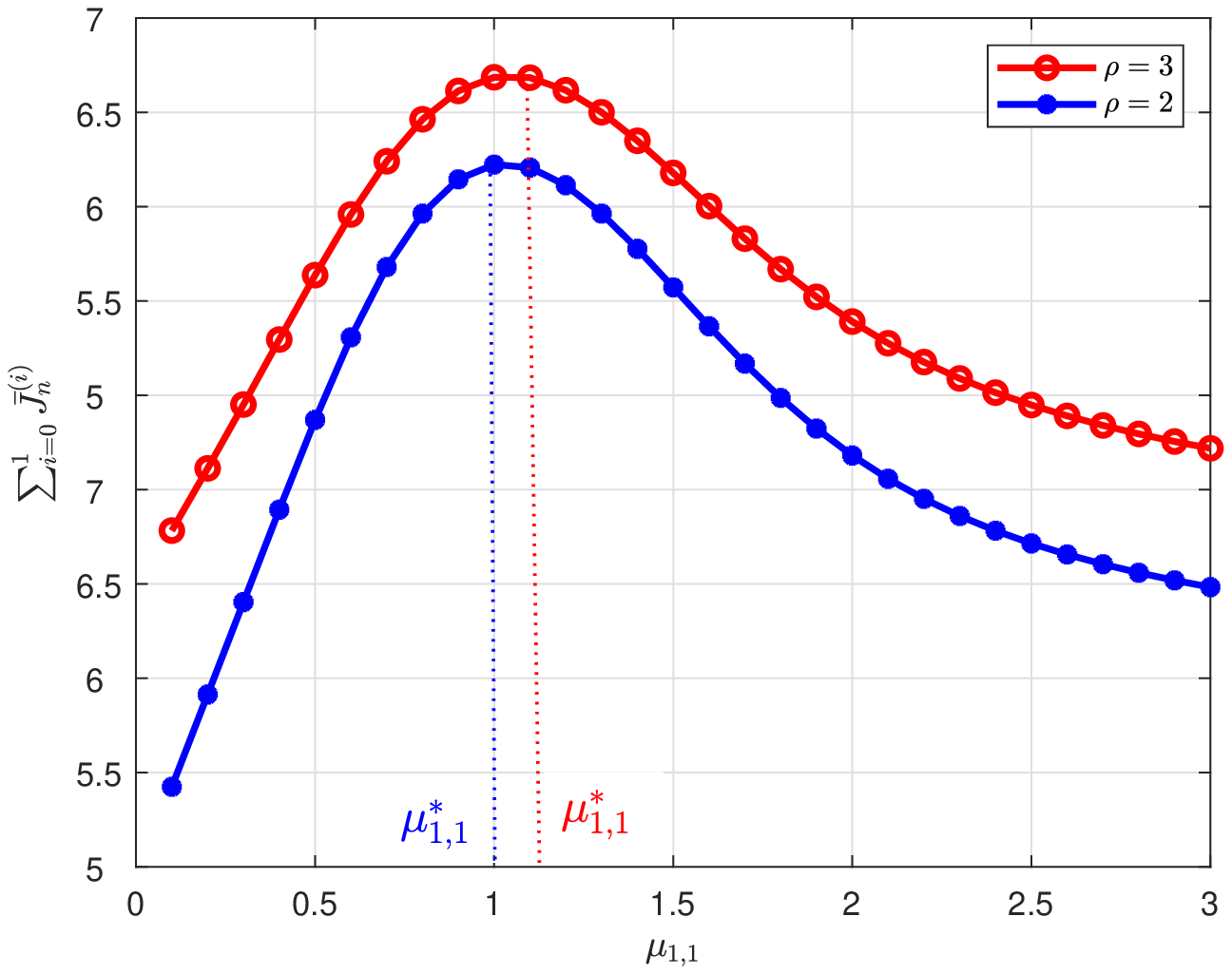}
 \caption{{ $c_{1,1}=0.5$ }}
\label{f7}
\end{subfigure}
\caption{ $K=5,~ c_{1,0}=0.3,~\gamma_{g_1}=1.$ 
}
\end{figure}
%
Considering one sensor and $L \!= \! 2$, the optimization variables are $\{c_{1,0}, c_{1,1}, \mu_{1,1}\}$. Fig. \ref{f6} illustrates the objective function $\sum_{i=0}^{L-1}\bar{J}_1^{(i)}$  versus the scale factor $c_{1,1}$.  We observe that the  objective function is not a concave function of $c_{1,1}$. Still there exists a point,  denoted as $c_{1,1}^*$,  at which the function attains its maximum. 
%
%
Starting from small values of $c_{1,1}$, as $c_{1,1}$ increases (until it
reaches $c_{1,1}^*$), the function value increases, because the harvested energy can recharge the battery and can yield more power for data transmission. However, when $c_{1,1}$ exceeds $c_{1,1}^*$, the harvested
and stored energy cannot support the data transmission and the function value decreases. 
Fig. \ref{f7}  shows  the objective function  versus the quantization threshold $\mu_{1,1}$. We observe that the  objective function  is not a
concave function of $\mu_{1,1}$.
Still there exists a point, denoted as $\mu_{1,1}^*$, at which the function achieves its maximum. 
%

%
\begin{table}[t]
\center
\scriptsize
\begin{tabular}{  |c|c||c c c|  } 
\hline
& &\makecell{\vspace{-2mm} \\ $\Pr(B_1=0)$ \\ }& \makecell{\vspace{-2mm}  \\$\Pr(B_1=50)$\\ } & \makecell{\vspace{-2mm}  \\$\bar{B}_1$\\ }\\
\hline\hline
(a)&\makecell{$\mu_{1,l}=[0,~ 0.8,~ 1.2,~ \infty]$\\$c_{1,l}=[0.3,~0.4,~0.2]$}& $\approx 0$ & 0.0451&31.97\\[.2cm]\hline
(b)&\makecell{$\mu_{1,l}=[0,~ 0.8,~ 1.2,~ \infty]$\\$c_{1,l}=[0.5,~0.7,~0.9]$}& 0.0318 & 0.0023 & 14.33\\[.2cm] \hline
(c)&\makecell{$\mu_{1,l}=[0,~ 0.1,~ 2,~ \infty]$\\$c_{1,l}=[0.4,~0.6,~0.3]$}& 0.0265 & 0.0039 & 15.32\\[.2cm] \hline
(d)&\makecell{$\mu_{1,l}=[0,~ 0.01,~ 0.1,~ \infty]$\\$c_{1,l}=[0.4,~0.6,~0.3]$}& $\approx 0 $ & 0.0357 & 28.32\\[.2cm]
\hline
\end{tabular}
\caption{The values of $\Pr(B_1\!=\!0),\Pr(B_1\!=\!50),\overline{B}_1$ for $K\!=\!50,~\rho\!=\!10,~\gamma_{g_1}\!=\!1$.}
\label{table1}
\end{table}
To accentuate the effect of the optimization variables on the entries of $\boldsymbol{\Phi}$ we define the average energy stored at the battery of sensor $n$ as $\overline{B}_n\!=\!\mathbb{E}\{B_n\}\!=\!\sum_{k=0}^K k~\phi_{n,k}$,
where the largest possible value for $\overline{B}_n$ is $K$. Table \ref{table1} shows $\Pr(B_1=0),~ \Pr(B_1=50), ~\overline{B}_1$  for four choices (a), (b), (c), (d).
%
%
%
Going from (a)  to (b), we note that given $\mu_l$'s, as $c_l$'s increase data transmit power in \eqref{alpha} increases. Due to large energy energy consumption for data transmission 
$\overline{B}_1$ decreases and the chance of energy outage increases.
%
%
Going from (c)  to (d), we note that given $c_l$'s, as  $\mu_l$'s decrease,  $\Pr(B_1=50)$ increases and $\Pr(B_1=0)$ decreases, and $\overline{B}_1$ increases.
Due to small energy consumption for data transmission, the chance of having near full battery increases,  
%
%
indicating that sensor has failed to utilize the excess energy. Both energy outage and energy overflow inevitably impact transmission and detection performance, leading to a reduction in the objective function.

$\bullet$ {\bf Accuracy of different search methods in solving (P2) and  detection performance-transmit power trade-off}: 
%
%
%
\begin{figure}[!t]
\centering
\includegraphics[width=70mm]{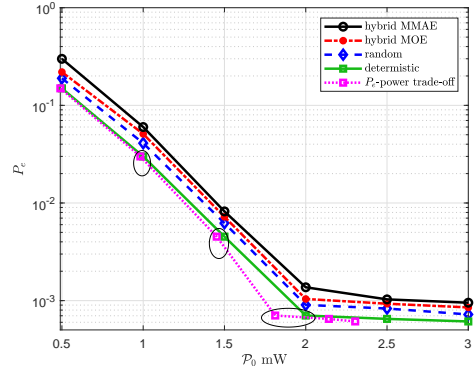}
 \caption{{ $P_e$ vs. ${\cal P}_0$  for $N=3,~K\! =\!5,~L=2,\rho\! =\! 2,~ \sigma^2_{w_n}\!=\! 1,~\gamma_{g_n}\! =\! 2,~P_{\text{d}_n}\! =\! 0.9, \forall n$, SNR$_s\! =\! 3$dB. }}
\label{f9}
\end{figure}
%
%
%
%
First, we compare the accuracy of deterministic, random, and hybrid search methods in Section \ref{how-to-solve-P1} in solving (P2).  
%
%
Fig.~\ref{f9} shows $P_e$ versus ${\cal P}_0$
for $L \!= \! 2$. To plot the curve  labeled as ``deterministic'' first we obtain the {\it optimal} solution, set transmit power control strategy in (\ref{alpha}) accordingly, and run Monte-Carlo simulation to find $P_e$. Similarly, we plot the curves labeled as ``random'', ``hybrid MMAE'', ``hybrid MOE'' using the {\it c-optimal} solution, the {\it sub-optimal} solution corresponding to MMAE, and the {\it sub-optimal} solution corresponding to MOE, respectively. When using RRS algorithm we choose the parameters of exploration phase $p\!=\!0.99, r \!=\! 0.1$, leading to $Q_1 \!=\!44$. For exploitation phase, we choose 
$Q_2 \!=\! 10$ for ``random''  and $Q_2 \!=\! 3$ for ``hybrid MMAE'' and ``hybrid MOE''.
%
%
%
%
%
Note that for all curves, as ${\cal P}_0$ increases $P_e$ decreases, which is expected. Also,  ``random'', ``hybrid MMAE'' and ``hybrid MOE'' perform very close to ``determistic''. 
%
%
Fig.~\ref{f9} also allows us to examine the existing trade-off between the average transmit power and the detection performance.   Consider the curve labeled ``$P_e$-power trade-off'' in Fig.~\ref{f9}, which shows how much average transmit power is required to provide a certain $P_e$ value. This curve is obtained from examining the points on  ``deterministic''  and  checking whether the power constraint in (P2) is active or inactive. At a given point, when this constraint is active (inactive), the average transmit power is equal to (less than) ${\cal P}_0$. Note that as ${\cal P}_0$ increases and $P_e$ reaches an error floor, the average transmit power is less than ${\cal P}_0$.
%

Since finding the {\it sub-optimal} solution has the lowest computational complexity, and its performance is very close to the {\it optimal} solution, from this point forward, we focus on ``hybrid MMAE'' and ``hybrid MOE''. 

\begin{figure}[!t]
 \begin{subfigure}[t]{0.24\textwidth}
 \centering
  \includegraphics[width=47mm]{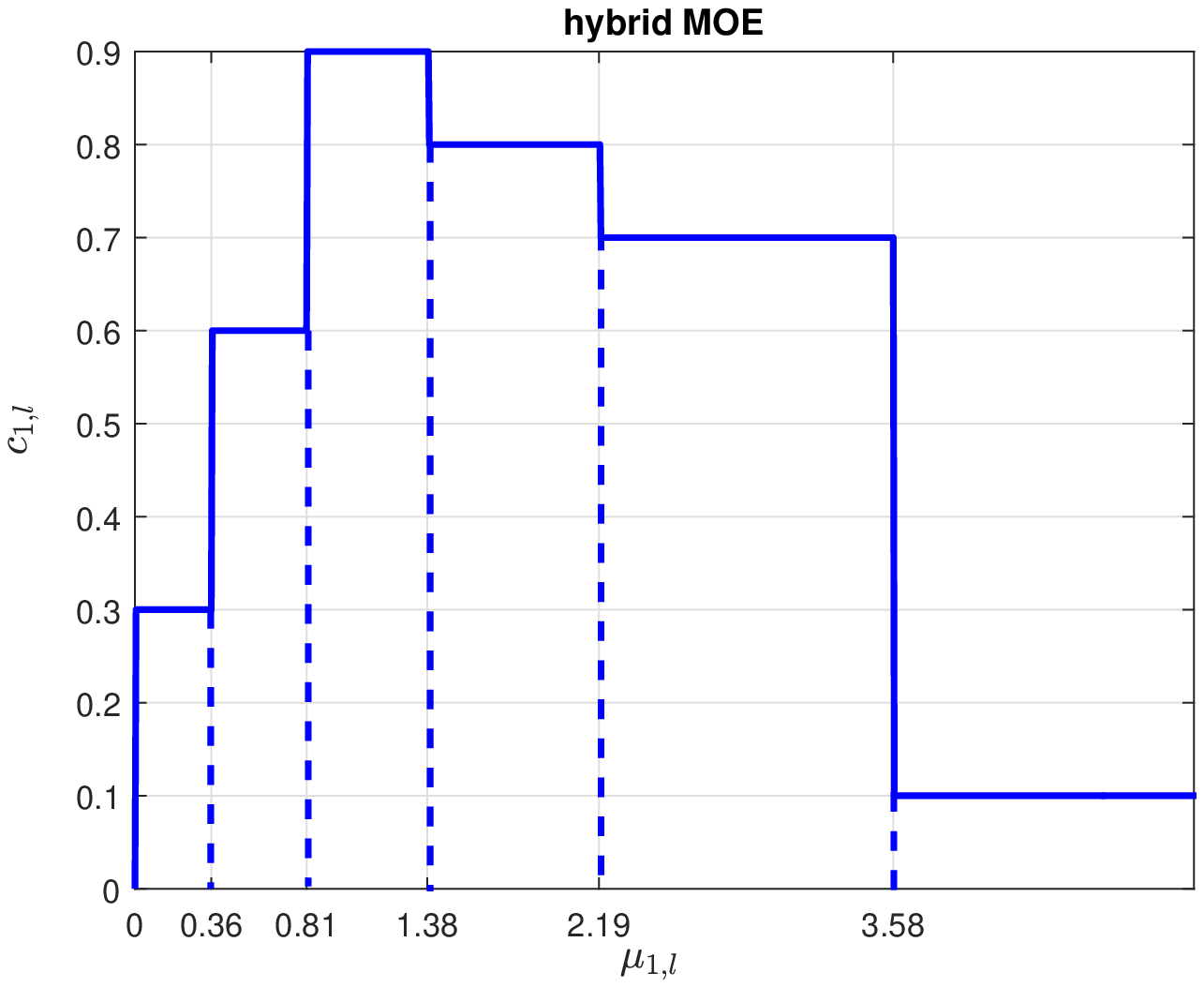}
 \caption{{ sub-optimal solution corresponding to hybrid MOE }}
\label{f16}
\end{subfigure}
\begin{subfigure}[t]{0.24\textwidth}
  \centering
  \includegraphics[width=47mm]{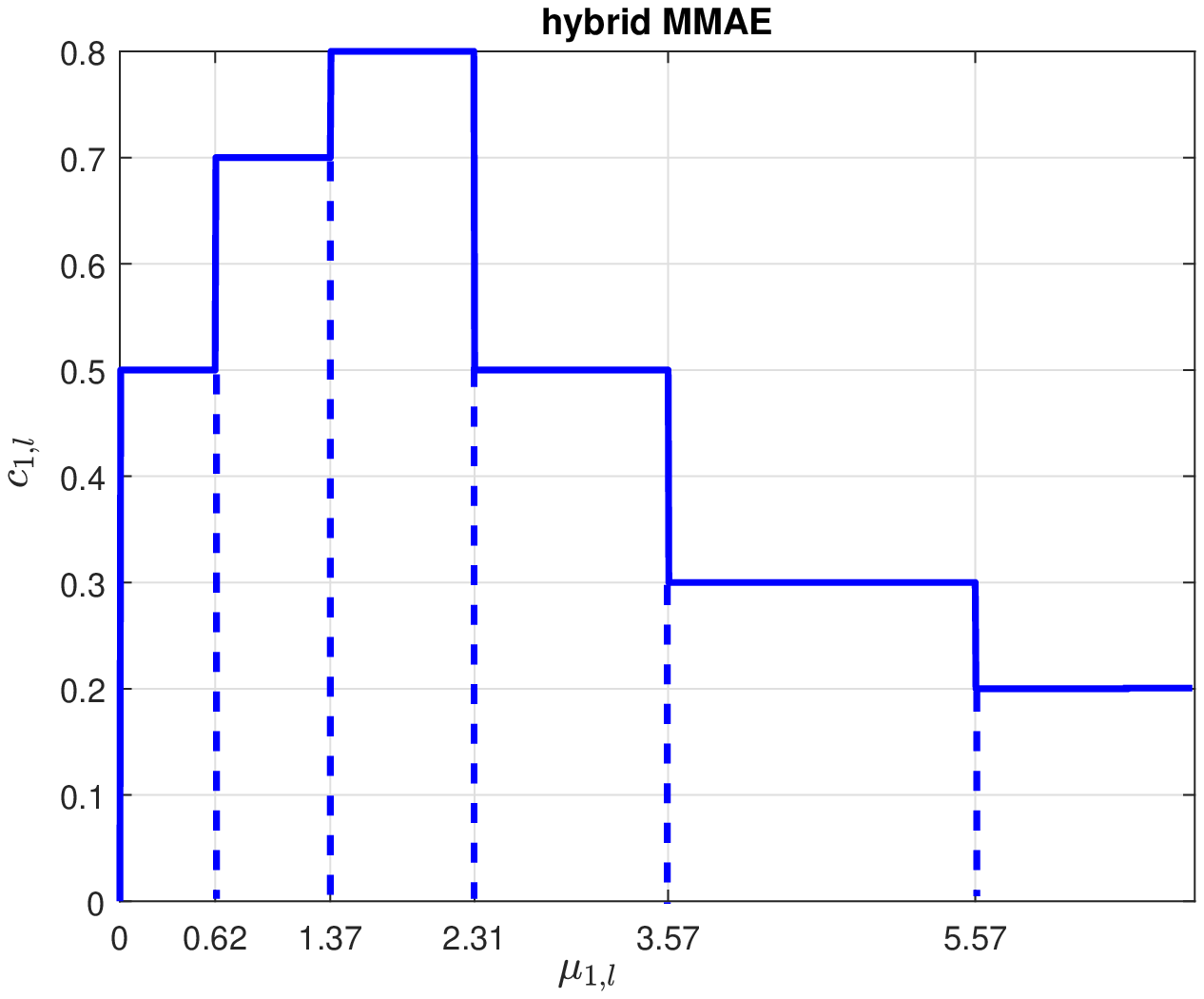}
 \caption{{  sub-optimal solution corresponding to hybrid MMAE}}
\label{f17}
\end{subfigure}
\caption{$K=5,~L\!=\!6,~\rho\!=\!2,~\sigma^2_{w_1}\!=\! 1,~ \gamma_{g_1}\!=\!2,~P_{\text{d}_1}\!=\!0.9,$ $\mathcal{P}_0\!=\!2$mW, SNR$_s\! =\!2$ dB.}
\end{figure}

$\bullet$ {\bf Behavior of the optimized  scale factors}:
Considering one sensor and $L\!=\!6$, the optimization variables are $\{c_{1,l}\}_{l=0}^5, \{\mu_{1,l}\}_{l=1}^5$. Fig. \ref{f16} and Fig. \ref{f17} depict the optimized $\{c_{1,l},\mu_{1,l}\}$'s corresponding to ``hybrid MMAE'' and ``hybrid MOE''. We note that, as $l$ increases (i.e., channel gain $g_{n,t}$ increases), the length of quantization interval $(\mu_{1,{l+1}}-\mu_{1,{l}})$ becomes larger. Also, $c_{1,l}$ first increases and then decreases. Considering (\ref{alpha}) this implies that, given the battery state $k$, as  $g_{n,t}$ increases ${\cal P}_{n,t}$ first increases and then decreases.

%

\begin{figure}[!t]
\centering
\includegraphics[width=70mm]{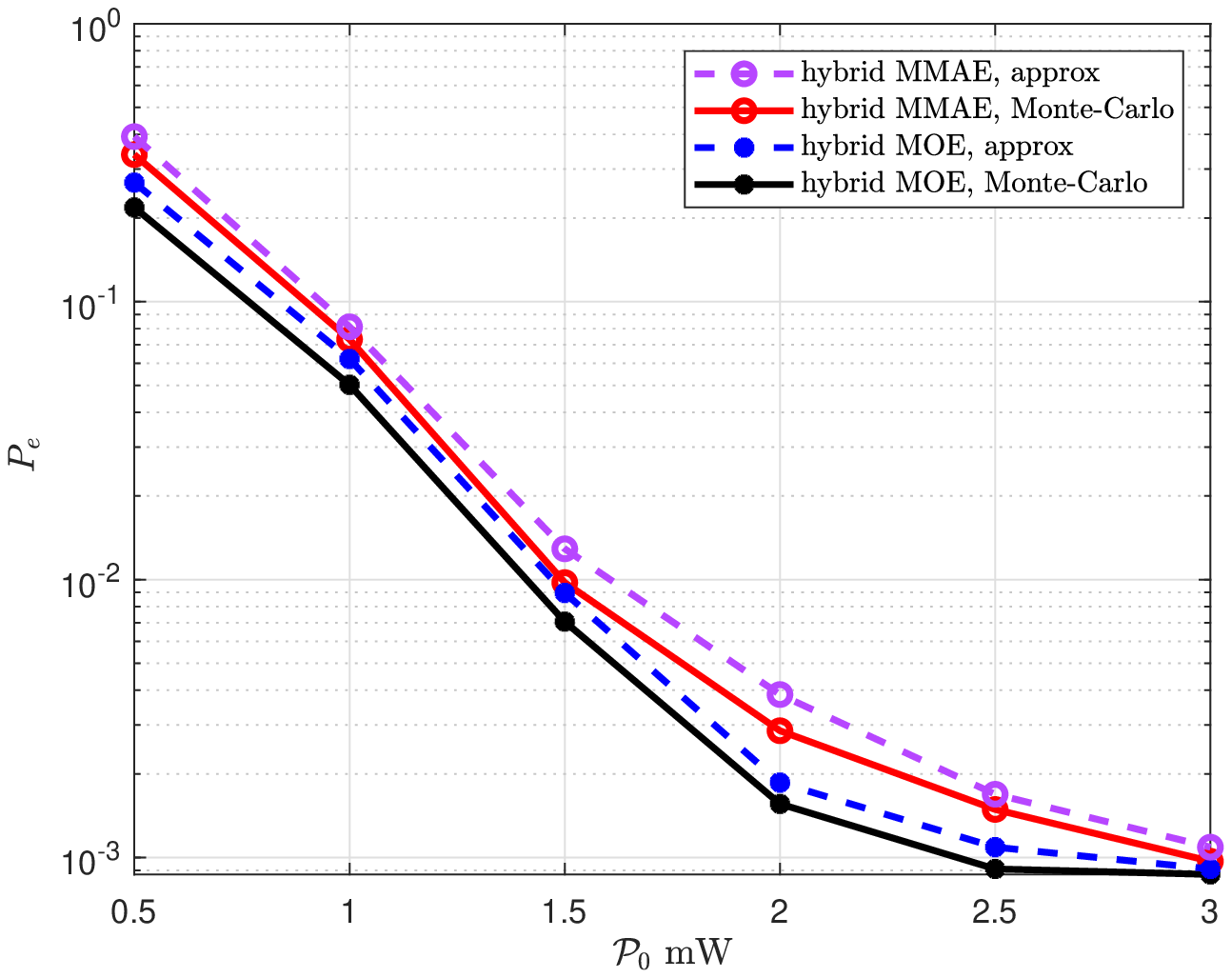}
 \caption{ $P_e$ vs. ${\cal P}_0$ for $N\!=\!5,~K\!=\!5,~L\! = \!3, \rho\!=\!2,~\sigma^2_{w_n}\!=\! 1~\gamma_{g_n}\!=\!2,~  P_{d_n}\!=\!0.9, \forall n$, SNR$_s=3$ dB.}
\label{f10}
\end{figure}

$\bullet$ {\bf Accuracy of $P_e$ approximate in  \eqref{clt_pe}}: 
To examine the accuracy of $P_e$ approximate in  \eqref{clt_pe}, we focus on 
 ``hybrid MMAE'' and ``hybrid MOE''.  Fig. \ref{f10} plots $P_e$ versus ${\cal P}_0$, in which $P_e$ values obtained from Monte-Carlo simulations are denoted as ``Monte-Carlo'', and $P_e$ values obtained from \eqref{clt_pe} are denoted as ``approx''.
 This figure suggests that the $P_e$ approximate in  \eqref{clt_pe} is reasonably accurate.   Henceforth, from this point forward, we use \eqref{clt_pe} to plot $P_e$.
%
%

$\bullet$ {\bf Dependency of $P_e$ on different parameters}:
%
\begin{figure}[!t]
\centering
\includegraphics[width=70mm]{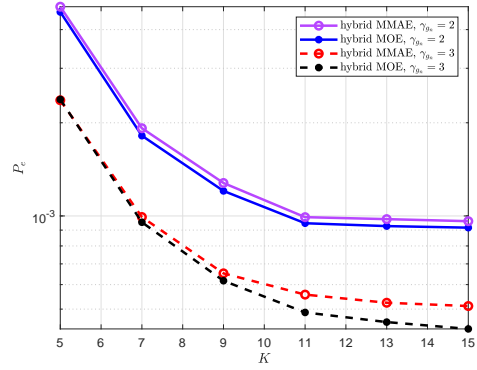}
 \caption{$P_e$ vs. $K$ for $N=5,~L=3,~\rho=5,~\sigma^2_{w_n}\!=\!1,~ P_{\text{d}_n}\!=\!0.9,\forall n, \mathcal{P}_0\!=\!3$mW, SNR$_s\! =\!5$dB.}
\label{f11}
\end{figure}
\begin{figure}[!t]
\centering
\includegraphics[width=70mm]{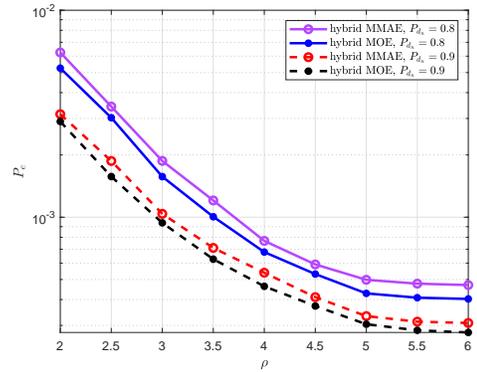}
 \caption{ $P_e$ vs. $\rho$ for $N\!=\!5,~K\!=\!5,~ L\!=\!3,~\sigma^2_{w_n}\!=\!1,~\gamma_{g_n}\!=\!3, \forall n, \mathcal{P}_0\!=\!3$mW, SNR$_s\! =\!3$ dB.}
\label{f12}
\end{figure}
\begin{figure}[!t]
\centering
\includegraphics[width=70mm]{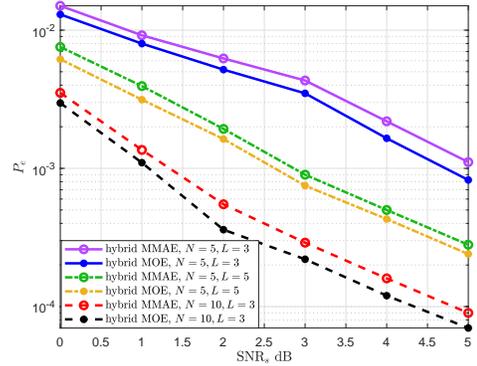}
 \caption{$P_e$ vs. SNR$_s$ for $K\!=\!5,~ \rho\!=\!2,~\sigma^2_{w_n}\!=\! 1,~ \gamma_{g_n}\!=\!2,~P_{\text{d}_n}\!=\!0.9,\forall n, \mathcal{P}_0\!=\!2$mW.}
\label{f14}
\end{figure}
%
Fig.~\ref{f11}-\ref{f14} plot $P_e$ corresponding to ``hybrid MMAE'' and ``hybrid MOE'' in terms of different system parameters. 
%
Fig. \ref{f11} depicts $P_e$ versus $K$ as $\gamma_{g_n}$ changes.  As $K$ increases $P_e$ decreases, until it reaches an error floor. 
This is because for large $K$, power ${\cal P}_{n,t}$ in (\ref{alpha}) is 
no longer restricted by $K$, and instead it is restricted by $\rho$.  Also, the communication channel noise $\sigma^2_{w_n}$ becomes dominant and leads to an error floor. Clearly, the error floor becomes smaller when $\gamma_{g_n}$ increases. Also, ``hybrid MOE'' outperforms ``hybrid MMAE''. 
%
%
%
%
Fig. \ref{f12} shows $P_e$ versus $\rho$ as  $\overline{P}_{\text{d}}$ changes. 
As $\rho$ increases $P_e$ decreases,  until it reaches an error floor.
This is because for large $\rho$, power ${\cal P}_{n,t}$ is no longer limited by the amount of harvested energy. Instead, $\sigma^2_{w_n}$ becomes the dominant factor and leads to an error floor. Also, increasing $\overline{P}_{\text{d}}$ lowers the error floor.
%
%
%
%
%
Fig. \ref{f14} shows $P_e$ versus SNR$_s$ as $N,L$ vary. Given $N,L$ as SNR$_s$ increases, $P_e$ decreases. Increasing $N$ and $L$ reduce $P_e$. Also, as $L$ increases, the gap between ``hybrid MMAE'' and ``hybrid MOE'' decreases. 
\section{Conclusions}\label{conclu}

We developed a power control strategy for an EH-enabled WSN, that is tasked with solving a binary distributed detection problem. Our proposed strategy is
parametrized  in  terms  of the channel gain quantization thresholds and the scale factors, which play key  roles  in  balancing  the  rates  of  energy  harvesting and energy consumption for transmission. We explored the optimal and sub-optimal strategies such that 
 the $J$-divergence based detection metric is  maximized,  subject to  an  average  transmit  power  per  sensor  constraint.  These optimization  problems  can  be  solved offline and allow each sensor to adapt its power based on its battery state and its quantized CSI (acquired via limited feedback from the FC). Since our non-convex optimization problem is not differentiable with respect to the optimization variables, we explored deterministic, random, and hybrid grid-based search methods, and showed that our proposed hybrid search methods have a low-computational complexity and  near-optimal performance. The structure of the optimized scale factors reveals that, given the battery state, the optimized power level is not a monotonic function of the channel gain.  We examined the existing trade-off between the average transmit  power  and  the  detection  performance.
We also demonstrated that increasing $K$ or $\rho$ do not necessarily lower the detection error, and it depends on the communication channel noise.
%
\bibliographystyle{IEEEtran}
\bibliography{RefEH}
\end{document}